\documentclass{cup-pan}
\usepackage{blindtext}
\usepackage{xcolor}
\usepackage{natbib}

\title{Updates to ALMA Site Properties:\\ using the ESO-Allegro Phase RMS database \\ ALMA Memo 624}

\author[1]{Luke T. Maud}
\author[2]{Andr\'es F. P\'erez-S\'anchez}
\author[3,4,5]{Yoshiharu Asaki}
\author[1]{Felix Stoehr}
\author[3]{Bill Dent}
\author[1]{Mar\'ia D\'iaz Trigo}

\affil[1]{ESO Headquarters, Karl-Schwarzchild-Str 2 85748 Garching, Germany. Email: \url{lmaud@eso.org}}
\affil[2]{Allergro - ALMA ARC node, Leiden Observatory, Leiden University, PO Box 9513.png, 2300 RA Leiden, The Netherlands}
\affil[3]{Joint ALMA Observatory, 
        Alonso de C\'{o}rdova 3107, Vitacura, Santiago, 763 0355, Chile}
\affil[4]{National Astronomical Observatory of Japan, 
        Alonso de C\'{o}rdova 3788, Office 61B, Vitacura, Santiago, Chile}
\affil[5]{National Astronomical Observatory of Japan, Osawa 2-21-1, Mitaka, Tokyo 181-8588, Japan}       
\corrauthor{Luke T. Maud}

\runningauthor{Maud et al.}

\begin{document}

\maketitle

\begin{abstract}
We present a long-term overview of the atmospheric phase stability at the Atacama Large Millimeter/submillimeter Array (ALMA) site, using $>$5\,years of data, that acts as the successor to the studies summarized two decades ago by \citet{Evans2003}. Importantly, we explore the atmospheric variations, the `phase RMS', and associated metadata of \textbf{over 17000} accrued ALMA observations taken since Cycle 3 (2015) by using the Bandpass calibrator source scans. We indicate the temporal phase RMS trends for average baseline lengths of 500, 1000, 5000, and 10000\,m, in contrast to the old stability studies that used a single 300\,m baseline phase monitor system. At the ALMA site, on the Chajnantor plateau, we report the diurnal variations and monthly changes in the phase RMS on ALMA relevant baselines lengths, measured directly from data, and we reaffirm such trends in atmospheric transmission (via Precipitable Water Vapour - PWV). We confirm that day observations have respectively higher phase RMS and PWV in contrast to night, while the monthly variations show Chilean winter (June - August) providing the best, high-frequency and long-baseline observing conditions - low (stable) phase RMS and low PWV. Yet, not all good phase stability condition occur when the PWV is low. Measurements of the phase RMS as a function of short timescales, 30 to 240\,s, that tie with typical target source scan times, and as a function of baseline length indicate that phase variations are smaller for short timescales and baselines and larger for longer timescales and baselines. We illustrate that fast-switching phase-referencing techniques, that allow short target scan times, could work well in reducing the phase RMS to suitable levels specifically for high-frequencies (Band 8, 9 and 10), long-baselines, and the two combined. Understanding and counteracting atmospheric variations is fundamental for interferometric synthesis arrays and we provide a forward look for improving observing scenarios to use the conditions at the ALMA site most optimally in order to maximize the science return. The Phase RMS database will be available along with helper scripts on the ALMA science portal for public use.




\end{abstract}

\section{Introduction}
\label{sec:intro}

Since the `Site Properties and Stringency' ALMA Memo 471 \citep{Evans2003} there has not been any global exploration of atmospheric phase stability, `phase RMS', conditions despite the many thousands of available ALMA datasets taken since observations began. The Precipitable Water Vapour (PWV) content has been explored in detail, as extracted from the APEX telescope\footnote{http://archive.eso.org/wdb/wdb/asm/meteo\_apex/form}, measurements at ALMA (using the Water Vapour Radiometer - WVR - system \citealt{Nikolic2013}) and other site measurements \citep{Otarola2019,Cortes2020} dating from 2006. However, for interferometric arrays the PWV parameterises directly only the transmission, and not the all important atmospheric phase stability which is \textit{fundamental} to being able to correlate signals from all antennas in an array successfully.   

There have been dedicated campaigns focused on e.g. long baseline conditions \citep{ALMA2015,Asaki2016,Matsushita2017}, and those looking to offer the high frequency observing capability at the longest baselines \citep{Asaki2020,Asaki2020b,Maud2020,Maud2022}. The latter studies specifically investigated the phase RMS conditions given the critical nature in conducting such `difficult' observations and the importance of phase stability. Globally a study about phase stability conditions for ALMA is lacking. ALMA does have a feedback loop in place where conditions are checked during observing and from previous observations. This is generally used in the context of the scheduling of observations, or to understand whether the observation currently being or already taken meet certain quality assurance (QA) parameters (See the ALMA Technical Handbook\footnote{https://almascience.eso.org/documents-and-tools}, and \citealt{Petry2020}). Table \ref{tab:octile} shows the nominal PWV levels with respect to associated highest observing bands, which are used as a reference for scheduling observations\footnote{The nominal PWV values are not necessarily hard-limits for the observing bands as the observing strategy can be adjusted when scheduling dependent on other factors such as antenna number, or longer time on source in higher (worse) PWV at a given band.}, while Table \ref{tab:phlim} shows the phase RMS, as path length variation in $\mu$m (see Section \ref{sec:rmsdb}), for a limit of 30$^{\circ}$. A phase RMS of 30$^{\circ}$ can be considered as a `very good' to `ideal' value (see also Section \ref{sec:impphase}). 


\begin{table}[!t]
\caption{ALMA PWV octiles, the nominal PWV values for each octile and associated highest observing bands.}
\label{tab:octile}
\begin{center} 
\begin{tabular}{l l l }
\headrow  \thead{Octile} & \thead{Nominal PWV (mm)} & \thead{Associated Bands} \\ 
 1 & 0.472   & Band 5$^a$, 9, 10 \\
 2 & 0.658   & Band 8$^b$, 9 \\
 3 & 0.913  &Band 7, 8$^c$ \\
 4 &1.262 & Band 7\\
 5 &1.796 & Band 5$^d$, 6\\
 6 &2.748 & Band 4\\
 7  &5.186  & Band 3 \\

\end{tabular}
\end{center}
\small{$^a$Band 5 around the 183\,GHz water line.\\
$^b$Band 8 at the upper end of the band, $>$430\,GHz.\\
$^c$Band 8 at the lower end of the band, $<$430\,GHz.\\
$^d$Band 5 away from the 183\,GHz water line.\\

}

\end{table}



The \textbf{ESO-Allegro Phase RMS database development study}, of which this ALMA memo presents the key results, was undertaken to make a retrospective systematic analysis of all 12\,m main array data taken from Cycle 3 to Cycle 7\footnote{The study is still ingesting data taken after 2019 and is planning to continue.}. Using the Bandpass source scan of each observation, a snapshot of the phase RMS conditions as a function of short timescales ($<$5-10\,min) and baseline length can be established from each observing run. Combining over 17000 EBs (including some from 2020 and 2021), standardised to certain baseline lengths and with measurements over a range of timescales, allows for an investigation of the \textit{actual} phase stability conditions at the ALMA site in recent history and how they vary diurnally and monthly, or correlate with other metadata (e.g. PWV, wind speed, temperature). The harvested phase RMS parameters can also be used to investigate how changing observation strategies for e.g. long-baselines and high-frequencies using fast-switching (or dynamic) phase referencing, could be utilized to optimally use the conditions at the ALMA site (Section \ref{sec:optim}), and also to remedy a common misconception often held that low PWV means conditions are always good and that successful high frequency observations can be made (Section \ref{sec:PWVph}).

\newpage
\section{Background}
\label{sec:back}
\subsection{Chajnantor Plateau and overview of the Evans 2003 Memo}
ALMA is situated on the Chajnantor plateau at an altitude of 5050\,m in the Andes of northern Chile. It is one of, if not the best, site for ground based sub-millimeter wavelength astronomy. The ALMA site is high and dry. The majority of studies on weather conditions or site parameters that discussed long-term (years) atmospheric phase stability, as also noted in the \citet{Evans2003} memo (using data from 1995 to 2002), are from pre-construction surveys and investigations that used ad-hoc monitoring systems at fixed locations on the plateau rather than tied with actual ALMA antenna positions. In short, these studies present both opacity/PWV data and phase RMS measurements using a dedicated short, 300\,m, baseline two element phase monitoring system\footnote{The ALMA memo series is where tens of phase RMS specific memos are stored: https://library.nrao.edu/alma.shtml}. In brief, diurnal and seasonal variations in opacity and phase RMS are observed \citep[e.g.][and references therein]{Holdaway1997,Radford2001,Butler2001}. In the summer\footnote{Throughout this memo, unless stated, we broadly use `summer' and 'winter' only related to December to March, and April to November, respectively.} months (starting December and through into March in Chile) there is the so-called `Altiplanic winter' \citep[see,][for a meteorological explanation]{Otarola2019} where opacity/PWV is significantly elevated as moist air is drawn over the Andes. Generally, conditions in and around February are so bad that ALMA does not observe and uses this month for maintenance tasks (see the ALMA Proposer's Guide\footnote{https://almascience.eso.org/proposing/proposers-guide}). The winter months (broadly April through to November, and specifically June, July and August) are those with notably lower opacity/PWV despite being modulated with diurnal variations, where day time has higher values than the night \citep[see also,][]{Otarola2019,Cortes2020}.

\begin{table}[bt]
\caption{ALMA bands, the frequency range of the bands, the frequency used to calculate the path length variations for a phase RMS of 30$^{\circ}$, and the path length variation in $\mu$m for the phase RMS of 30$^{\circ}$.}
\label{tab:phlim}
\begin{center}
    \begin{tabular}{l l l l}
\headrow \thead{Band} & \thead{Frequency Range} & \thead{Frequency } & \thead{Path Length Variation ($\mu$m)}   \\ 
\headrow \thead{} & \thead{(GHz)} & \thead{(GHz)} & \thead{for  30$^{\circ}$ phase RMS}  \\ 

Band 10 & 787-950 & 850 &29       \\
Band 9&  602-720  & 650  &38     \\
Band 8 &  385-500  &  440  &57    \\
Band 7   & 275-373  & 320   &78 \\ 
Band 6   & 211-275  &  240  &104 \\ 
Band 5  & 163-211  &   185  &135 \\  
Band 4 &  125-163  &  140  &178 \\  
Band 3   & 84-116   &  100    &250 \\  
Band 1$^*$  & 35-50    &  40      &624   \\ 
\end{tabular}
\end{center}
\small{$^*$Band 1 is offered for ALMA Cycle 10.}
\end{table}

Similarly, the phase monitor system showed that the phase RMS on a 300\,m baseline follows closely the pattern of sun rise and sun set, in that $\sim$00:00 to 13:00 UTC (night) exhibits the lowest phase RMS, while a clear hump begins to emerge at $\sim$13:00 UTC and peaks around 19:00 UTC (day/afternoon), see \citet[][Figure 5]{Evans2003}. As \citet{Evans2003} note, the phase RMS from their 10\,min data samples indicates that the atmosphere is only stable enough to conduct short baseline or low frequency observations for the majority of the time without any phase compensation. They indicate that only 10\,\% of time could be used for e.g. $\sim$500\,GHz observations (ALMA Band 8, with a 30$^{\circ}$ phase RMS limit). Although discussed, the statistics presented did not incorporate phase compensation, i.e. phase referencing or a WVR system, which act to lower the phase RMS and would allow higher frequency and longer baseline observations for a larger proportion of time. 

\subsection{The Phase RMS database}
\label{sec:rmsdb}

For every observation execution block (EB) observed on the 12\,m main array listed as a Quality Assurance level 2 (QA2) semi-pass or pass we obtain the raw (ASDM) data of the Bandpass calibrator source along with the WVR information. The ASDMs are imported to Measurement Sets as per typical data reduction using {\sc casa} \citep{Casa2022}. We extract and store metadata\footnote{The PWV is always reported with respect to Zenith, while all metadata are extracted for the entire observing time, not just the Bandpass scan length.}, such as PWV, wind speed, temperature and also extract information about the calibrators and targets, such as source names and elevation. All parameters are stored in an SQL database which is described further in the Appendix. WVR phase corrections are applied to all datasets using the {\sc wvrgcal} command, such that we have two parallel phase-time streams of data, with and without WVR correction. Subsequently, the Bandpass phase solutions are obtained on an integration timescale (typically between 3-6\,s for the ALMA 12\,m main array observations). Integration based solutions are possible because the Bandpass calibrators are all point-like strong sources with high signal-to-noise ratios (SNR). Using the {\sc gaincal} command in this process also means that automatic flags are produced for any bad or corrupted integration(s) or antenna(s). From the antenna based phase gaintables the baseline based phase-time streams are reconstructed and the phase RMS values are then calculated (for all baselines) and as a function of time. 

For all EBs we calculate the phase RMS for timescales of 30\,s, 45\,s, then 60\,s to 240\,s, in 30\,s steps which is possible given that the Bandpass observations are at least 5\,min long. The minimum of 30\,s is chosen as this relates closely with realistic target scan times that could be used at ALMA (see also Section \ref{sec:ALMAobs}). Specifically, we extract the phases for numerous overlapping samples of a given timescale from the phase-time stream as we move from the start to the end of the data by one integration, shifting the sampling window each time, while calculating the phase RMS\footnote{In practice we use the standard deviation which avoids having to make an additional solution to offset each phase-time stream to an average zero-degrees phase, as is required when calculating the RMS. The RMS, when using a sample with zero mean is exactly the same as the standard deviation.} per sampled window. The \textit{final} phase RMS stored, for a given baseline and timescale, is the average from all of the sampling windows. In recent works the phase RMS measured as a function of timescale in this manner is shown to be a reasonable proxy for establishing the amount of phase RMS that likely remains in the science target source scans after phase correction using phase-referencing on that timescale \citep{Maud2020,Maud2022}\footnote{There is much discussion in the various Holdaway ALMA memos that the timescale should be half the phase referencing cycle time. However in \citet{Maud2022} this would result in overly optimistic values of phase RMS from their ALMA observations, while \citet{Carilli1999} note that the factor 2 is appropriate for self-calibrated data where the phase solution is at the mid-point of the timescale of a target scan.}. We also store the residual phase RMS which are measured \textit{after} the application of an ideal phase gain solution made on a given timescale, and the so-called D-phase, that is the difference between the phase gain solutions themselves (see the Appendix for further details). These are not used nor discussed in this memo.

Importantly, \textbf{the phase RMS is stored as a frequency independent parameter\footnote{If the EB was taken in Band 3 (100\,GHz) and achieved 10$^{\circ}$ phase RMS, this is equivalent to $\sim$83\,$\mu$m, which, for example then relates to a Band 6 (250\,GHz) phase RMS of $\sim$25$^{\circ}$.}, `Path Length' variation in units of microns}, irrespective of the observing band (frequency) used. The path length variation ($\ell$, in $\mu$m) can be established from $\sigma_{\phi}$ (radians), the phase RMS of the phase, $\phi$, for a given observing frequency ($\nu_{obs}$, in Hz) via:

\begin{equation}
  \label{eqn1}
 \ell  =  \frac{\sigma_{\phi}.c}{2\pi .\nu_{obs}}\quad(\mu m),
\end{equation}

\noindent where $c$ is the speed of light (in $\mu$m\,s$^{-1}$)\footnote{Excluding possible atmospheric dispersion effects around deep atmospheric absorption lines that can alter the path length within a given band \citep[see][although subsequent, unpublished, ALMA testing found these result could be a factor of 2 worse than in reality]{Curtis2009}.}. Our sample is \textbf{not} limited only to the best conditions, in terms of phase RMS, just because the conditions in some cases allowed higher frequency bands to be used. The variations in path length for a given observation are simply a property of the atmospheric fluctuations at that time. Some low frequency observations were made in conditions that could have been used for higher frequencies, especially when there were no high frequency projects at a given time during the day or night. As we report the phase RMS in microns we can always calculate what the highest possible observing frequency could have been. The caveat of our sample is that \textit{we are} limited to `better' conditions, in terms of PWV and likely phase RMS, as we use data that was suitable for ALMA to observe in. For example, ALMA does not observe in February when the PWV can exceed $>$10\,mm (see Section \ref{pwvsite}) and hence we clearly miss such high PWVs in our sample as compared to continual monitoring \citep{Otarola2019,Cortes2020}. We have no phase RMS measures from February to compare with. It is not critical that we miss the very worst conditions as ALMA would not be able to observe, although it must be remembered that our presented statistics are therefore percentages or fractions based upon the time when ALMA \textit{can} observe, and are not referenced to time available in a calendar year.   

\subsubsection{Making a Comparable Phase RMS Sample}
\label{sec:comparable}
\begin{figure}[!h]
\centering
\includegraphics[width=0.75\textwidth]{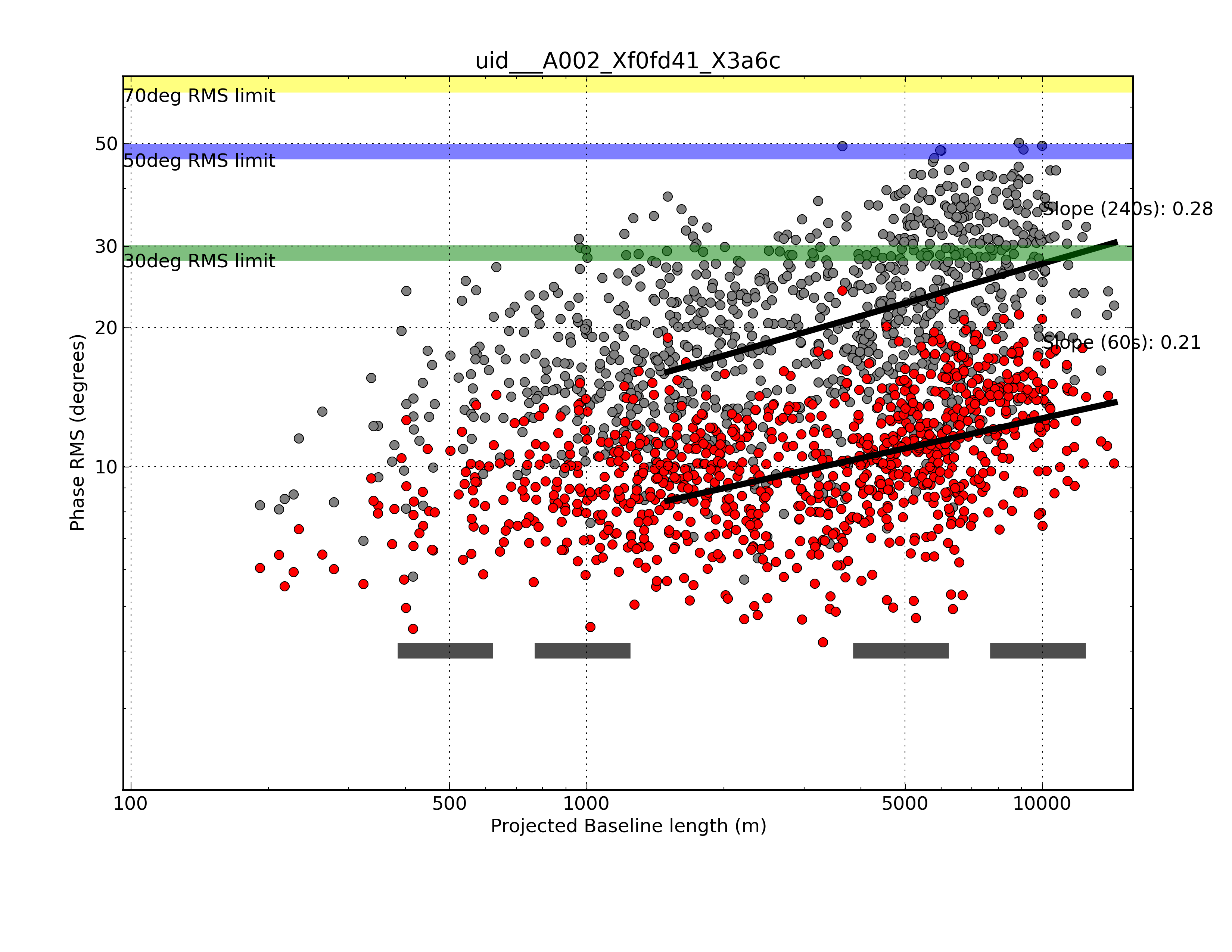}
\caption{Typical plot of the phase RMS after WVR corrections are applied (in degrees) as calculated on 240\,s (grey) and 60\,s (red) timescales against baseline length for the Bandpass source scan of a long baseline observation. The slopes, as fitted using baselines $>$1500\,m, indicate that the 60\,s timescale is shallower than the 240\,s timescale (that matches closely \citealt{Matsushita2017}). Hypothetically, if this observation did not have baselines $>$5000\,m then scaling the 60\,s, 5000\,m phase RMS to 10000\,m using the \citet{Matsushita2017} slope of 0.29 then we would overestimate the phase RMS by $\sim$5\,\% as compared to the fitted slope. In reality we would not know the slope for a short baseline observation. The bars at the bottom of the plot indicate the ranges used for establishing the `Summary' baseline phase RMS values.} 
\label{fig:ssfslope}
\end{figure}

As ALMA is a reconfigurable interferometer there is a broad spread in baseline length between `Compact' and 'Extended' arrays with $<$500\,m and $>$10\,km maximal baseline lengths. We therefore produce `Summary' phase RMS values for 500, 1000, 5000 and 10000\,m baseline lengths (at all timescales). These are chosen to maximise the overlap with baselines in various ALMA array configurations. If these baseline lengths exist in a given EB (see also Figure \ref{fig:ssfslope}), then an average phase RMS from baseline lengths about the summary value are calculated and stored ($\pm$100, $\pm$200, $\pm$1000, $\pm$2000\,m for the four baseline lengths, respectively). In cases where any of the four summary baseline lengths do not exist in the EB, an extrapolation is made from the summary values that do exist\footnote{In the case of the shortest baseline configuration ($<$500\,m longest baseline) we scale to the 500\,m Summary value using the phase RMS calculated from the longest 20 percent of baselines.}. We use the scaling values from the long baseline Spatial Structure Function (SSF) investigations of \citet{Matsushita2017} following $b^{0.65}$ and $b^{0.22}$ for data without WVR correction applied, and $b^{0.60}$ and $b^{0.29}$ for data with WVR correction\footnote{Using WVR corrected data we scale the phase RMS value at the 500\,m summary baseline length by $\sim$1.51 to establish the value at 1000\,m, and thereafter we scale the 1000\,m value by $\sim$1.59 to calculate the 5000\,m value, and finally we use a factor of $\sim$1.22 to scale from 5000\, to 10000\,m.} for baselines, $b$, $<$1\,km and those $>$1\,km, respectively. There are 66\,\%, 48\,\%, 12\,\% and 8\,\% of the EBs with baselines covering 500, 1000, 5000 and 10000\,m respectively. We identify a possible caveat here that for longer baseline observations and short timescales, $<$60-90\,s, the phase RMS \textit{can} flatten from $\sim$1000-5000\,m (Figure \ref{fig:ssfslope}, \citealt[see also Figures 9 and 10 in][]{Asaki2020}, and \citealt[Figure 9 in][]{Maud2022}) and thus the baseline based scaling from shorter summary baselines could \textbf{overestimate} the phase RMS when projecting out to $\geq$5000\,m. 

\begin{figure}[bt]
\centering
\includegraphics[width=0.75\textwidth]{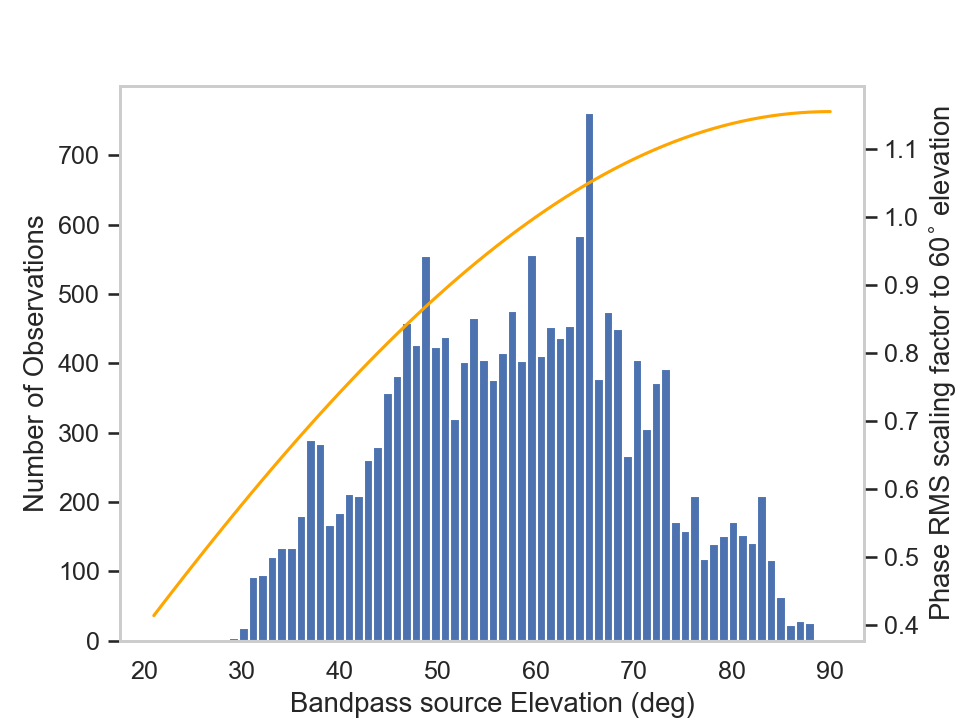}
\caption{Histogram of the Bandpass source elevations. The orange line indicates the scaling factor that would be needed to scale the phase RMS at the plotted elevations to a normalized 60$^{\circ}$ elevation. The phase RMS of low elevation observations would decrease, while those at high elevation would marginally increase. There are only 15 observations below 30$^{\circ}$ elevation which cannot be seen on the plot due to the y-axis scale.} 
\label{fig:elplot}
\end{figure}

The Bandpass source observations from which the phase RMS are calculated cover a range of observing elevations, from 21$^{\circ}$ to 88$^{\circ}$, where the median elevation of all observations is 58$^{\circ}$ (there are only 15 observations with elevation $<$30$^{\circ}$). Figure \ref{fig:elplot} shows the distribution of the Bandpass source elevations as well as the line indicative of the scaling factor that would be required to normalize the phase RMS values to an elevation of 60$^{\circ}$ (using the ratio of sin(elevation)/sin(60$^{\circ}$), see \citealt{Holdaway1995b}). If we were to scale everything to a 60$^{\circ}$ elevation, there are 3711 observations with an elevation $<$47$^{\circ}$ where the phase RMS would be \textit{reduced} by more than 15\,\%, while there are only 93 observations $>$85$^{\circ}$ elevation where the phase RMS would be \textit{increased} by $>$15\,\%. Considering this, the phase RMS values as directly measured from the data are higher, on average, than those we would obtain if we were to scale everything to a 60$^{\circ}$ elevation. Therefore, throughout the memo we present the phase RMS as measured under the premise that the distribution of elevations provides a more realistic representation of how science sources are observed on the sky at varying elevations.  

\newpage
\subsubsection{Summary of Main Points}
\begin{itemize}
    \item The database stores various metadata about each EB along with phase RMS values for a range of timescales for each baseline of the EB, and as established on summary baseline lengths of 500, 1000, 5000 and 10000\,m.
    \item The phase RMS are recorded and presented as a frequency (Band) independent \textit{path length variation} in units of microns. These values are irrespective of the band the observation itself was conducted in.
    \item The phase RMS as used throughout the report are the averages of the phase RMS values taken over a range of baseline lengths and from multiple overlapping time samples of the phase-time stream for each of those baselines using the Bandpass source scan.
    \item Extrapolating the short-baseline phase RMS values to longer baselines ($\geq$5000\,m) using the fixed baseline based power-law scaling could act to overestimate the phase RMS for shorter timescales ($<$60-90\,s).
    \item We miss the very worst conditions where observations, even at Band 3, could not be successfully made. In particular we are likely to report lower PWV values (at quartiles, percentiles and cumulatively) as compared to historic continual measurements.
    \item All percentages or fractions presented are relative to time in which ALMA can observe.
\end{itemize}

\begin{figure}[!h]
\centering
\includegraphics[width=0.70\textwidth]{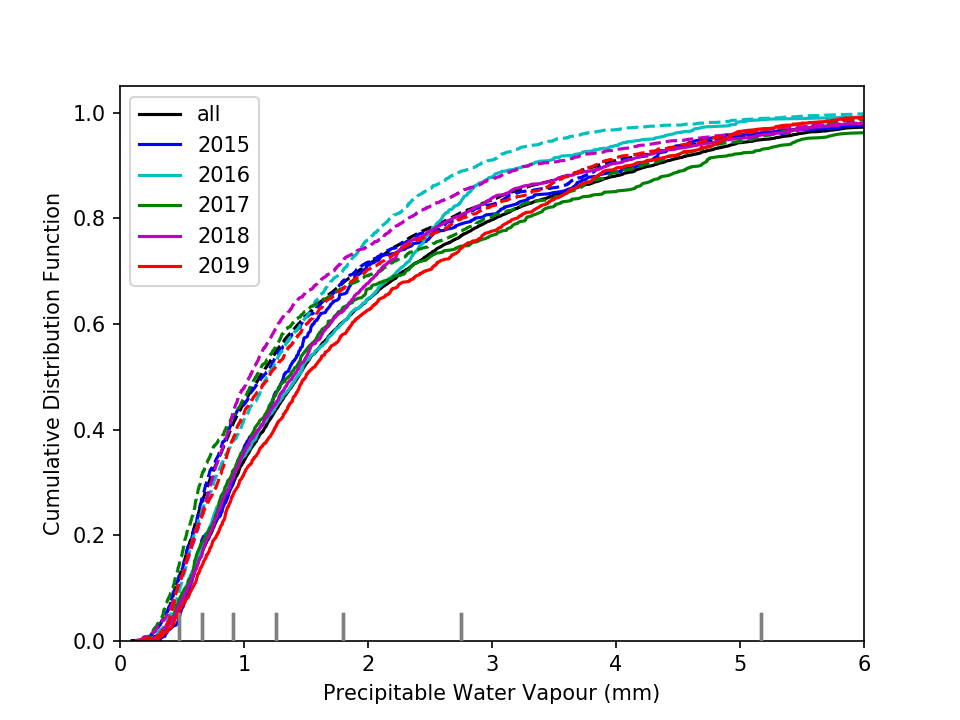}
\caption{Cumulative distribution function for day (solid) and night (dashed) PWV measurements from all data of years 2015 to 2019 inclusive (colours). Night time provides more lower PWV time for higher frequency bands to the order of $\sim$10\,\%. The grey markers indicate the various nominal PWV levels (Table \ref{tab:octile}).} 
\label{fig:pwvcdf}
\end{figure}

\section{Results}
\subsection{Precipitable Water Vapour}
\label{pwvsite}
\subsubsection{Overall Results}
For completeness we briefly report on the changes in PWV, but for further detailed analyses we direct the reader to \citet{Otarola2019,Cortes2020}. The PWV level directly translates to the transmission at ALMA and is used to determine which observing band can be used. Figure \ref{fig:pwvcdf} shows the cumulative distribution function for the PWV of the data ingested from years 2015 to 2019 inclusive (colours) for day (solid line) and night (dashed line). In all years the night time always offers lower PWV conditions overall. Comparing the top octiles, corresponding to nominal PWV values of $<$0.472, 0.658, 0.913 and 1.262\,mm PWV respectively, night time offers 5, 10, 11 and 11\,\% more `time' than the day (averaged over all years and months). From the conditions of the observations taken since Cycle 3, we indicate that the Band 7 nominal PWV level was met roughly $\sim$50-55\,\% of the time during the night, and $\sim$40-45\,\% of the time during the day. Observations with nominal Band 3 PWV conditions or better occurred $>$95\% of the time when ALMA observed on the sky. Year by year variations are of a few percent, but we do not have an extensive sample as compared to \citet{Otarola2019,Cortes2020}. We also reiterate that the reported percentages are based on the observations made, i.e. none during the high-PWV month of February. 


\begin{figure}[!h]
\centering
\includegraphics[width=1.0\textwidth]{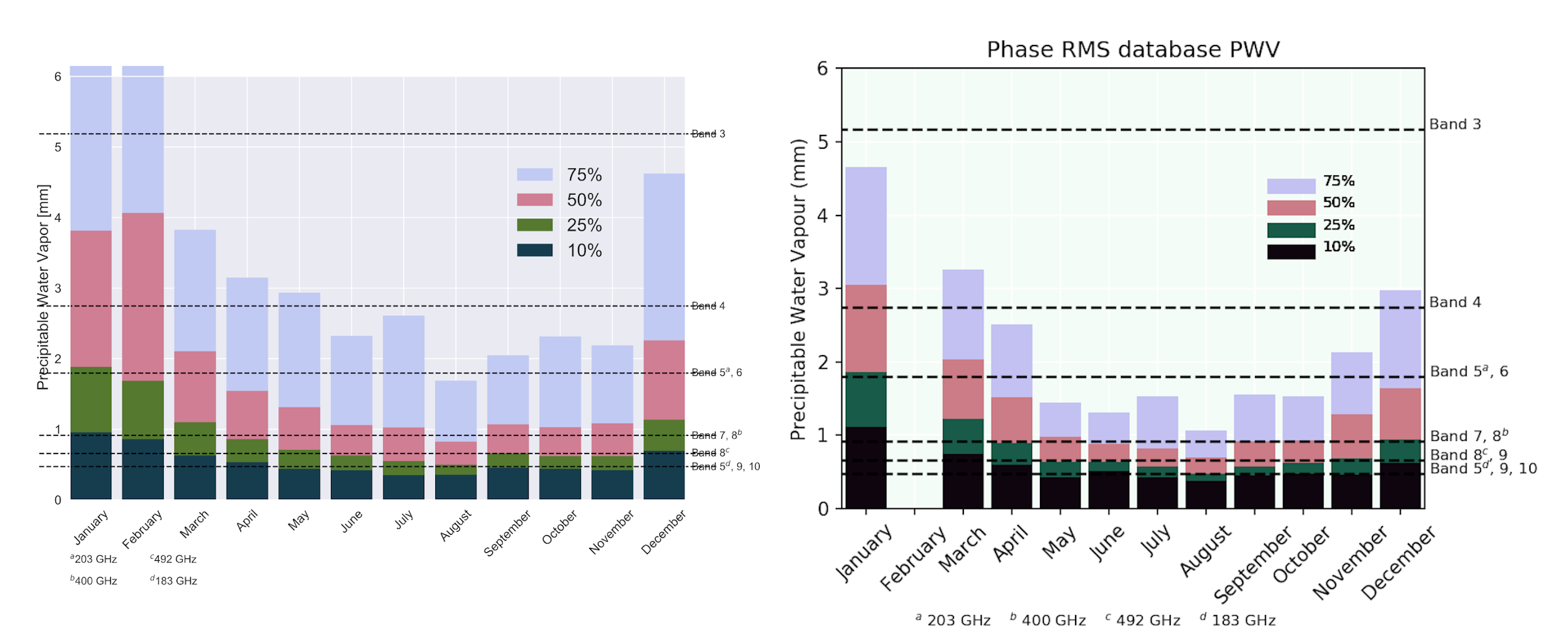}
\caption{Reproduction of the ALMA Proposer Guide Figure 2 (left) and our version as replicated using the PWV metadata stored in the Phase RMS database (right). At low PWV percentiles of 10\,\% and 25\,\% we indicate similar values, however our sample is biased to the conditions ALMA observed in and hence do not include the `worst' PWV. The horizontal lines indicate the nominal PWV values for different bands.} 
\label{fig:pwvqrt}
\end{figure}

\begin{table}[!h]
\caption{PWV level (in mm) for all months at 10, 25, 50 and 75\,th percentiles. February is excluded as there are no data.}
\label{tab:pwvpercent1}
\centering
\begin{tabular}{l l l l l l l l l l l l}
\headrow \thead{Percentile} & \thead{Jan.} & \thead{Mar.} & \thead{Apr.} & \thead{May} & \thead{June} &  \thead{July} &  \thead{Aug.} & \thead{Sep.} & \thead{Oct.} &  \thead{Nov.} & \thead{Dec.} \\    
75\,\% & 4.65 & 3.27  & 2.52 &  1.45 &  1.30  &  1.53 &  1.06 & 1.56 & 1.54  &  2.14 &  2.97  \\
50\,\% & 3.05  &  2.03 &  1.51 &  0.98 &  0.88 &  0.82 &   0.70 & 0. 92 &  0.94 &  1.29 & 1.64 \\
25\,\% &  1.86  & 1.23 &  0.90  &  0.64 &  0.64 &  0.58 &  0.48 & 0. 58 &  0.63 &  0.69 &  0.94 \\
10\,\% &  1.12  &  0.75 &   0.60 &  0.43 &  0.51 &  0.43 &   0.37 &  0.45 &  0.47 & 0.46 & 0.62 \\
\end{tabular}

\end{table}

\subsubsection{Seasonal Variation}
Figure \ref{fig:pwvqrt} (left panel) is a reproduction of Figure 2 from the ALMA Proposer's Guide that presents the PWV percentiles of 10, 25, 50 and 75\,\% as a function of months. The right panel is a replicated version using the PWV metadata from the Phase RMS database, excluding the month of February. 
Comparing to the Proposer's Guide figure, we show similar trends and similar values at the 10\,\% and 25\,\% levels, however, as we are limited to conditions when ALMA observed we report lower PWV values at the 50\,\% and 75\,\% levels, i.e. we indicate a larger fraction of `better' conditions. Our replicated figure is more representative in indicating the percentiles for data that are taken and therefore more tied with the conditions that the users can expect. However, the global view, as shown in the ALMA Proposer's Guide is the best representation in terms of total time and hence for planning how many hours ALMA could observe. In the months from May through to November the 50th percentile level indicates Band 8 nominal PWV conditions (Table \ref{tab:pwvpercent1}). 

\begin{figure}[!bt]
\centering
\includegraphics[width=0.95\textwidth]{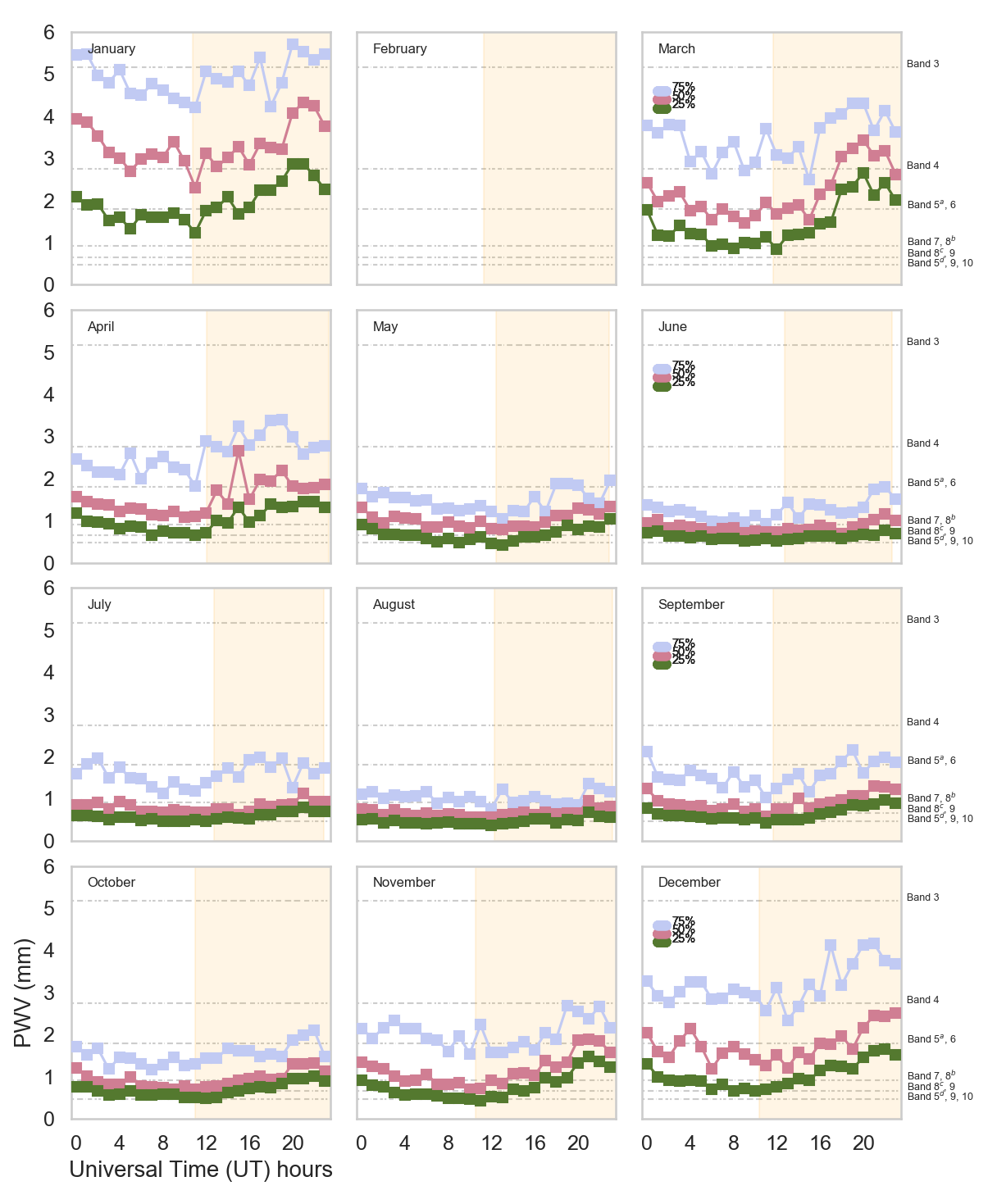}
\caption{PWV at 25, 50 and 75 percentile levels per month over a 24 hour period averaged over all years. The summer months, particularly December, January and March are the worst, while the winter months June, July, August offer $\sim$1\,mm conditions approximately 50\,\% of the time. The time axis is shown in UTC while the yellow band indicates sun-up and sun-set for the various months.}
\label{fig:pwv24}
\end{figure}

\subsubsection{Diurnal Changes}
Over a 24 hour timescale the 25, 50 and 75 percentiles do not fluctuate significantly (Figure \ref{fig:pwv24}). On average the maximum to minimum PWV level changes by a factor of 2.2, 1.9 and 1.8 respectively at each quartile. The day time generally shows an increase in PWV from $\sim$11-13 UT while the night time hours have the lowest PWV. During the winter the increase in PWV during day time is lower than that occurring in the summer months. 

\begin{figure}[!h]
\centering
\includegraphics[width=1.0\textwidth]{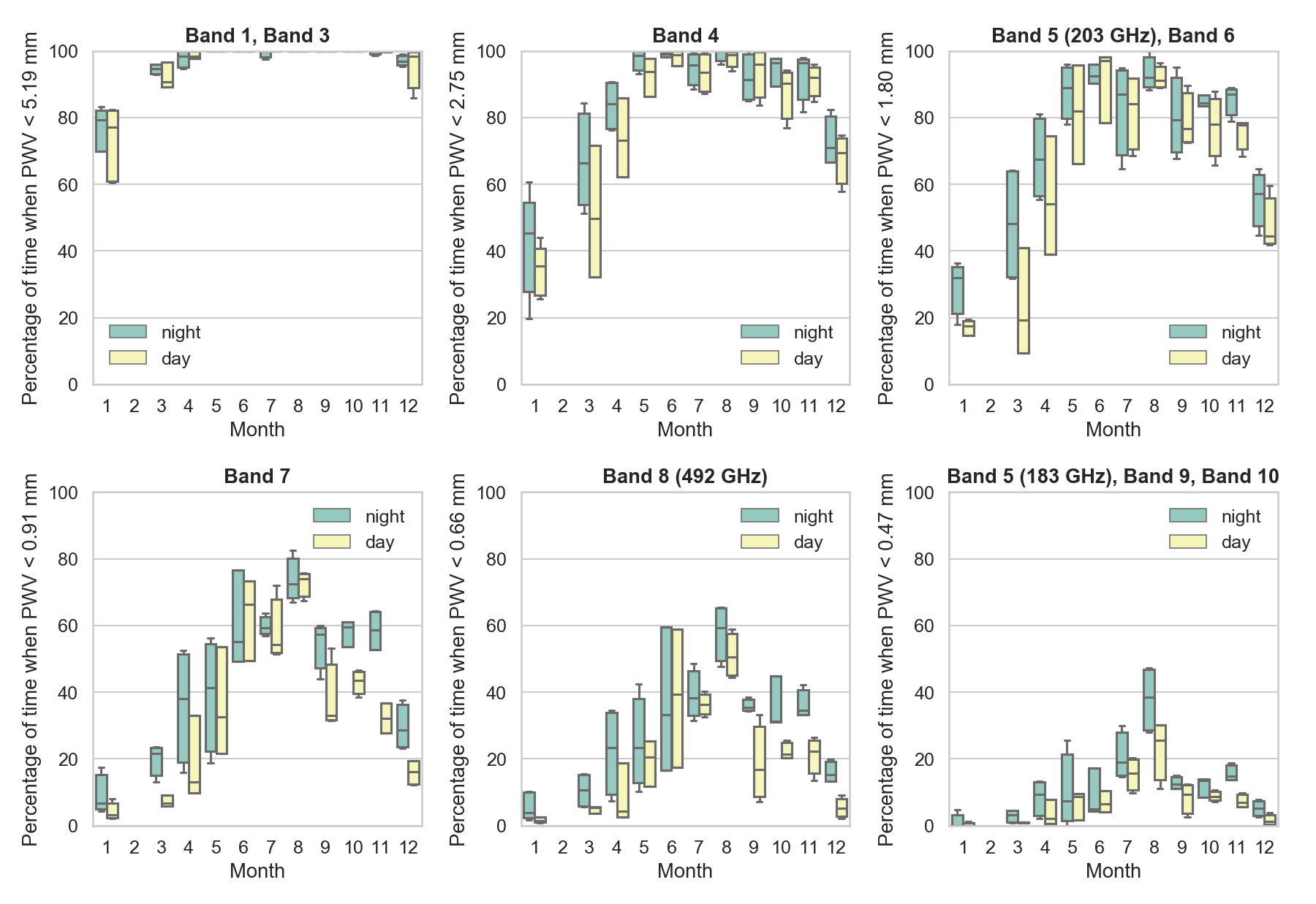}  
\caption{The percentage of time available for various bands using PWV from the Phase RMS database presented in the style of Figure 3 from the ALMA Proposer's Guide. The box-spread are 25 and 75\,\% while the whiskers are the 9 and 95\,\% levels considering the time available per year as measured using the observations since 2015. Day time uses the UTC range 1100-2300, otherwise observations are associated with night time. Note that Band 5 at 183\,GHz is paired with Band 9 and 10 as it requires very good transmission, while Band 1 can be conducted in Band 3 conditions that define the time availability in that panel.} 
\label{fig:fig3rep}
\end{figure}

\begin{table}[!ht]
\caption{Median percentage of time available (based on ALMA time on-sky) for each month, during day/night with cuts at the nominal PWV levels, 5.19, 2.75, 1.80, 0.91, 0.66 and 0.47\,mm}
\label{tab:pwvpercent}
\centering
\footnotesize
\begin{tabular}{l l l l l l l l l l l l}
\headrow \thead{PWV} & \thead{Jan.} & \thead{Mar.} & \thead{Apr.} & \thead{May} & \thead{June} &  \thead{July} &  \thead{Aug.} & \thead{Sep.} & \thead{Oct.} &  \thead{Nov.} & \thead{Dec.} \\    
5.19& 77/79 & 91/95 &98/98 &100/100&100/100&100/100&100/100&100/100&100/100&100/100&98/97  \\
2.75& 35/45  &50/66 & 73/84 & 94/99  & 99/99 & 93/95  &99/100 & 96/91 & 90/96&  92/96 & 69/70 \\
1.80 & 18/32 &  19/48 & 54/67 & 82/89 & 97/92 & 84/87 &  91/92&  77/79 & 78/84 & 78/87 &  44/57\\
0.91 & 3/7 & 7/22 & 13/38&  23/41 & 66/55&  54/59  &74/73 & 33/57& 43/59  &32/59 & 16/29 \\
0.66 & 2/4 & 5/11&  4/23  &21/23&  39/33& 36/38 & 50/59 & 17/35&  21/31 & 22/35  &5/15\\
0.47 & 0/0 & 1/3 & 2/9&  9/7& 6/5&  16/19&  26/38 & 9/12  &9/14 &7/14& 1/5 \\

\end{tabular}

\end{table}

\subsubsection{Impact on observing}

Figure \ref{fig:fig3rep} shows the day (we use 1100 to 2300 UTC, or 7\,am to 7\,pm where Chilean local time = UTC - 4) and night time percentages of time available where the PWV falls below the nominal PWV octiles for the associated bands, as shown in each panel, in the style of Figure 3 from the ALMA proposer guide. For the conditions in which ALMA made observations, Band 3 (and therefore Band 1) could have been conducted essentially at all times. 
For all bands there is the clear trend for both day and night in that the summer months November to April have the lowest percentage of `low' PWV time. The winter months June through to September provide the `best' low PWV conditions, primarily in the night, although the day median time available typically indicate only 5-10\,\% less time. It is clear that higher frequency observations are most possible in winter with Band 10 nominal PWV conditions averaging about 17\,\% of the time from July to October inclusive (day and night averaged). Table \ref{tab:pwvpercent} lists the median percentages of time available.




\subsection{Phase RMS}
\label{sec:phasemain}
\subsubsection{Overall Results}
\label{sec:overallph}
The Phase RMS is a main driver of the accuracy of any interferometric observation and can be the main limiting factor. The phase RMS must to be low enough during the scans of a particular science target after calibration as to allow accurate and reliable image reconstruction (see also Section \ref{sec:impphase}). As outlined in Section \ref{sec:rmsdb}, unlike previous studies \citep[e.g.][]{Evans2003, Holdaway1995, Butler2001}, we measure the phase RMS of ALMA data on a range of timescales that are relevant to phase referencing cycle times, essentially providing a proxy for the phase RMS that would \textit{remain} in the scans of a target \citep[see also][]{Maud2022} after phase calibration, and for ALMA relevant baseline lengths. ALMA observations follow different observing strategies using a range of cycle times for changing array configurations and observing bands that we discuss further in Section \ref{sec:ALMAobs}, here we provide the overall view of the measured phase RMS parameters for comparisons.

\begin{figure}[!ht]
\centering
\includegraphics[width=1.0\textwidth]{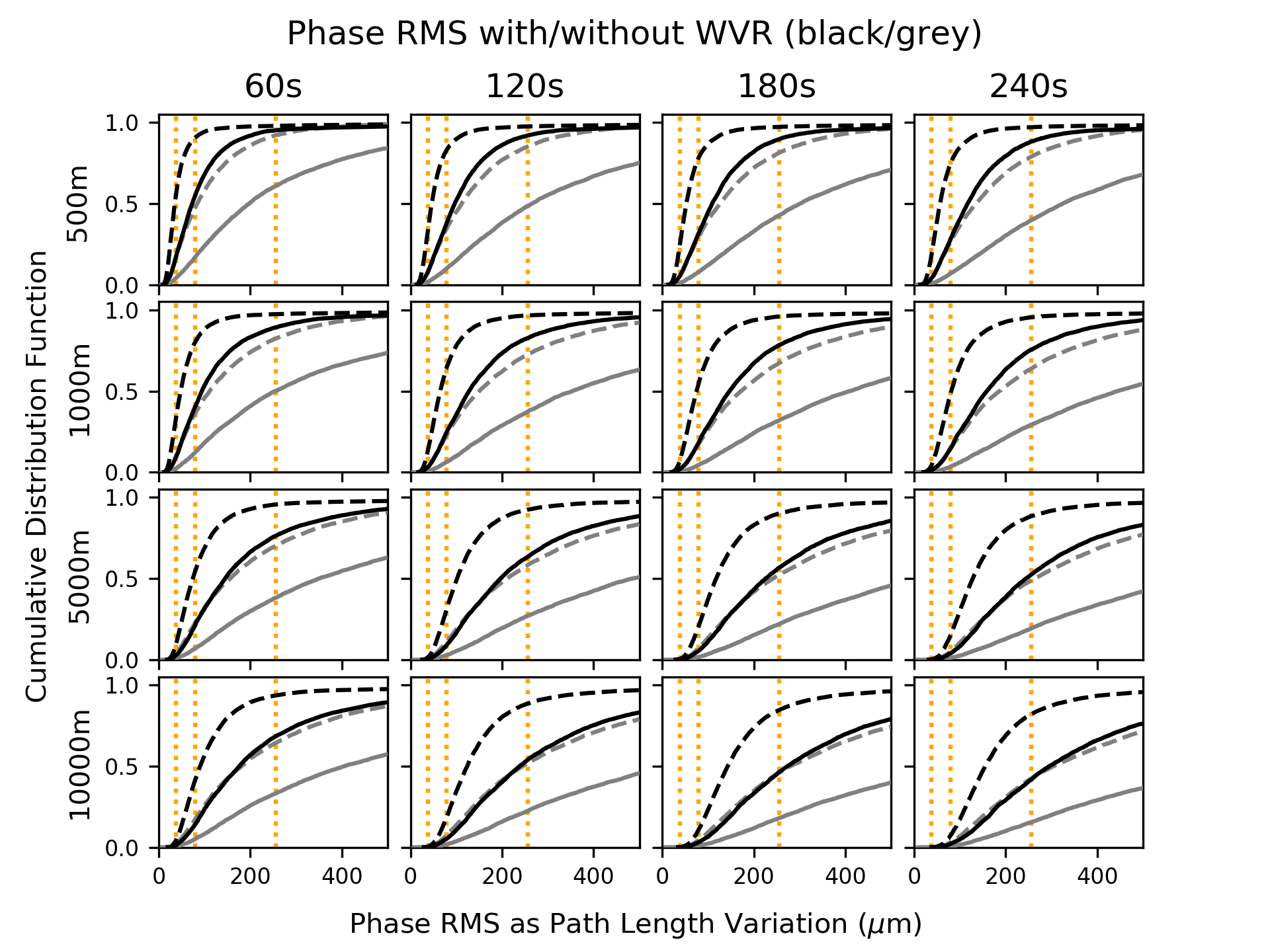}
\caption{Cumulative distribution functions for the phase RMS `Summary' values with increasing timescale moving left to right (60, 120, 180, 240\,s) and increasing baseline length moving top to bottom (500, 1000, 5000, 10000\,m). Black and grey indicate the with and without WVR corrections applied, while the dashed and solid lines are night and day (1100-2300 UTC). For all baselines and timescales the night time with WVR correction provides the largest amount of low phase RMS conditions. The orange dotted lines indicate the levels of 30$^{\circ}$ phase RMS (as a path length variation in $\mu$m) for Band 9, 7 and 3 from left to right in each panel.} 
\label{fig:phcum}
\end{figure}

Figure \ref{fig:phcum} shows the night (dashed) and day (solid) cumulative distribution functions combining all years of data using the phase RMS summary values for baseline lengths 500, 1000, 5000, 10000\,m (top to bottom) and timescales 60, 120, 180, 240\,s (left to right). The black and grey represent the phase RMS as measured with and without the WVR corrections applied, respectively, while the orange lines indicate the Band 9, 7 and 3 phase RMS of 30$^{\circ}$ (as a path length variation). The WVR system significantly improves the phase RMS and, interestingly, when applied pushes the day time observing to lower phase RMS as compared to night time conditions without the WVR corrections. As the WVR corrections are significant, and they generally improve all observations, most subsequent phase RMS plots and discussion relates to WVR applied values (see Section \ref{sec:wvrcorr} for WVR statistics). Overall, the shorter baselines and short timescales provide more low phase RMS conditions, while long baseline and long timescales \textit{do not} provide low enough phase RMS for the highest frequency bands (9 and 10).  

\begin{figure}[!h]
\centering
\includegraphics[width=0.75\textwidth]{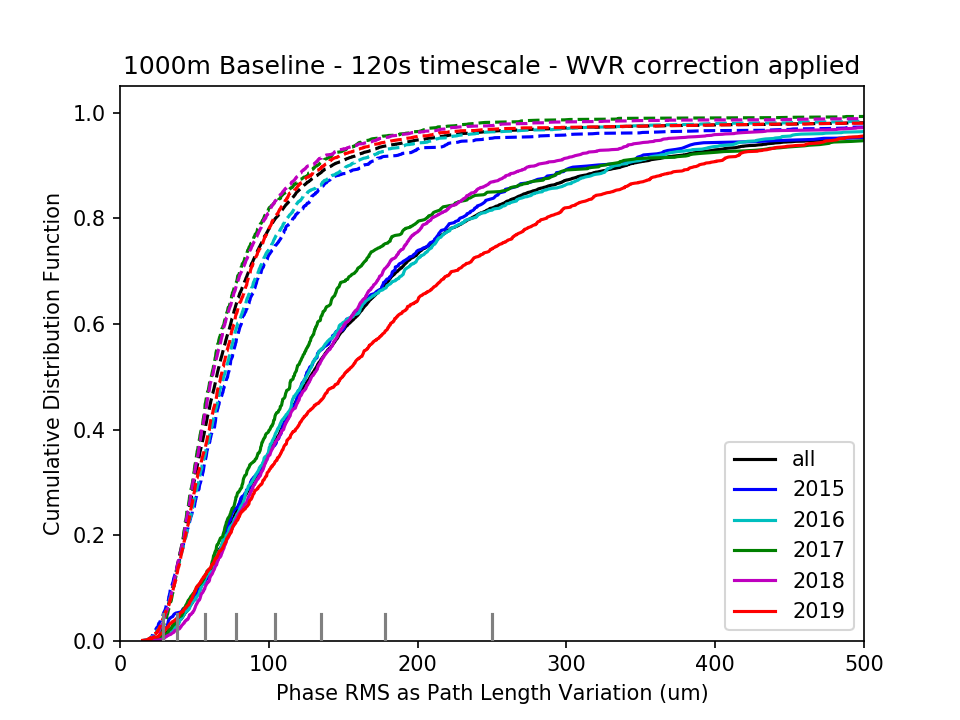}
\caption{Cumulative distribution function for the phase RMS at 120\,s on the 1000\,m summary baseline with WVR corrections applied. The dashed and solid lines are night and day (1100-2300 UTC) while the colours indicate the different complete years 2015 to 2019. The grey markers indicate the 30$^{\circ}$ phase RMS as a path length variation for the various bands (Table \ref{tab:phlim}). Generally at night there is $\sim$10\,\% variation in different years, and thus $\pm$5\,\% as compared with the cumulative function for all years. During the day the spread is roughly doubled. Overall the same trends hold for the other baselines and timescales.} 
\label{fig:phcum1000m}
\end{figure}

For short baselines there is less differentiation between timescales as compared to the longer baselines. For the 500\,m baseline length at night time with WVR correction the 30$^{\circ}$ Band 7 limit (see Table \ref{tab:phlim}) is achieved 74, 78, 83 and 90\,\% of the time for moving from the longest to shortest timescales, 240, 180, 120 and 60\,s. For the longest 10000\,m baselines there is almost a factor of 6 increase in the amount of time, 7, 11, 19, 41\,\%, that conditions meet the same band 7 limit moving from 240\,s to 60\,s (see also Section \ref{sec:ALMAobs}). This finding is tied with the atmospheric structure, in that the spatially large turbulent atmospheric cells providing the most phase variations are common to both antennas on short baselines. Interferometers are sensitive to turbulent cells of the order (of a few) of the scale of the baseline lengths as they move over the array. For short baselines the phase RMS saturates at timescales of a few minutes as many small-scale cells pass over the antennas creating the maximum variation whereas the spatially larger cells remain common to the antenna pairs and are not `seen'. The phase RMS can only be notably reduced for short baselines if moving to very short timescales, although this is pragmatically inefficient \citep[cf.][]{Masson1994} and short baselines anyway have intrinsically better phase RMS when measured on the same timescales as compared to longer baselines (as shown in Figure \ref{fig:phcum}). On the contrary, longer baselines are sensitive to the spatially larger turbulent cells and so the phase RMS increases as a function of time as many of the increasingly larger cells pass over the array that are not common to either antenna of a given baseline. Saturation of the phase RMS takes some tens of minutes to hours until the largest cells have completely crossed the array \citep[cf.][]{Matsushita2017}. The positive aspect is that in moving to shorter timescales (10s of seconds) there can be significant improvements (reduction) in the phase RMS for long baselines.


\begin{figure}[!h]
\centering
\includegraphics[width=0.95\textwidth]{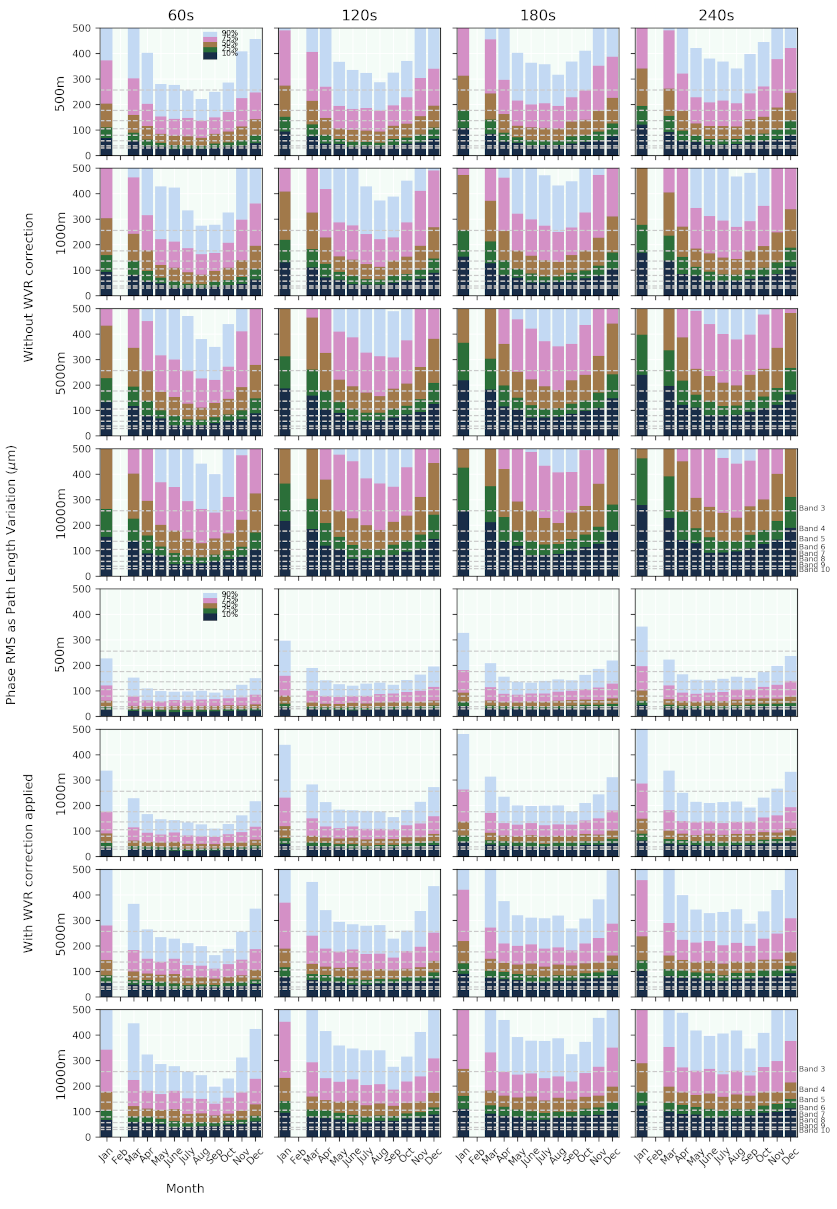}  
\caption{Phase RMS levels before WVR correction (upper) and after WVR corrections (lower), shown as path length variations in $\mu$m at the 10, 25, 50, 75 and 90 percentiles as a function of month for the various baseline and timescales. Moving left to right and top to bottom shows timescales of 60, 120, 180, 240\,s and baseline lengths of 500, 1000, 5000, 10000\,m respectively. Winter months provide more stable conditions, as do shorter baselines and shorter timescales.} 
\label{fig:phQs}
\end{figure}

At higher frequencies the lowest phase RMS, or path length variations, are required. Even on the shortest 500\,m baseline, at night with WVR applied, the cumulative distribution is increasing rapidly below $\sim$100\,$\mu$m. At the 30$^{\circ}$ Band 9 limit (38\,$\mu$m) there is notable difference when moving to shorter times, from 240 to 60\,s, where conditions meet the limit for 18, 24, 34 and 56\,\% of the time. For the longest baselines, again at night with WVR applied, there is $\sim$4.5\,\% of time available with a low enough phase RMS at the 60\,s timescale to meet the Band 9 limit, and effectively zero time available for longer timescales (also see Section \ref{sec:ALMAobs}). On the other hand, the longest baselines at night with WVR applied can be conducted in Band 3 $>$80\,\% of the time as it has the least strict phase RMS requirement except Band 1 which can essentially be conducted 100\,\% of the time at night, irrespective of the timescale (60\,s to 240\,s).  

In Figure \ref{fig:phcum1000m} we present the yearly cumulative distribution for the summary 1000\,m baseline and 120\,s timescale with WVR correction applied. Overall each year of data shows the same general trend with $\pm$5\,\% variation from the all year average at night, and $\pm$10\,\% for the day. The 500\,m baseline spread is reduced by a factor of $\sim$2, the 5000\,m baseline is very similar to the 1000\,s baseline as shown, while the 10000\,m baseline has a spread larger by a factor of $\sim$2.  

\subsubsection{Seasonal Variation}
\label{sec:phseason}

The above cumulative values provide a rather pessimistic view on the possibility of long baseline and high frequency observations, one of ALMA's key capabilities as outlined since before construction \citep{Richer2005,Wootten2009}. In Figure \ref{fig:phQs} we show the phase RMS, as path length variations, before (upper panels) and after WVR application (lower panels) at the 10, 25, 50, 75 and 90 percentile levels as a function of month in each panel, while left to right is for the 60, 120, 180 and 240\,s timescale and the top to bottom shows baseline lengths of 500, 1000, 5000 and 10000\,m (separately for observations before and after WVR correction).
Globally the winter months (June, July, August) provide the greatest proportion of low phase RMS conditions, although the increase of the various percentiles is typically a factor of 2 moving to summer months ($>$3-4 for phase RMS without WVR correction). Considering the months that ALMA can observe in, January is by far the worst month in terms of phase stability where the percentiles change is $>$2-3 as compared to winter months for WVR corrected phase RMS. For Solar mode observations at ALMA the WVR system cannot be used and so the significant yearly variations in phase RMS without WVR corrections must be accounted for in scheduling. Note that any diurnal variations are averaged out here. 


Comparing with PWV only, as shown in Figure \ref{fig:pwvqrt}, the nominal Band 8 ($<$0.913\,mm) condition is achievable in terms of transmission at the $\sim$50\,\% level from May through to October, but in terms of phase RMS (after WVR application) the short baselines indicate Band 8 conditions at the 50-75\,\% level while the longest baselines show Band 8 conditions consistent with only 10-25\,\%, depending on the timescale used. For the longer baselines the shortest timescales provide a notably larger fraction of observations with low Phase RMS conditions, of the order 2-3 times more. The PWV alone is too optimistic in estimating when the `most difficult' observations at higher frequencies and longer baselines could be taken, but on the contrary it is the limiting factor for shorter baselines where the phase RMS limits can be met more easily. 




\begin{figure}[bt]
\begin{minipage}{0.46\textwidth}
\includegraphics[width=1.2\textwidth]{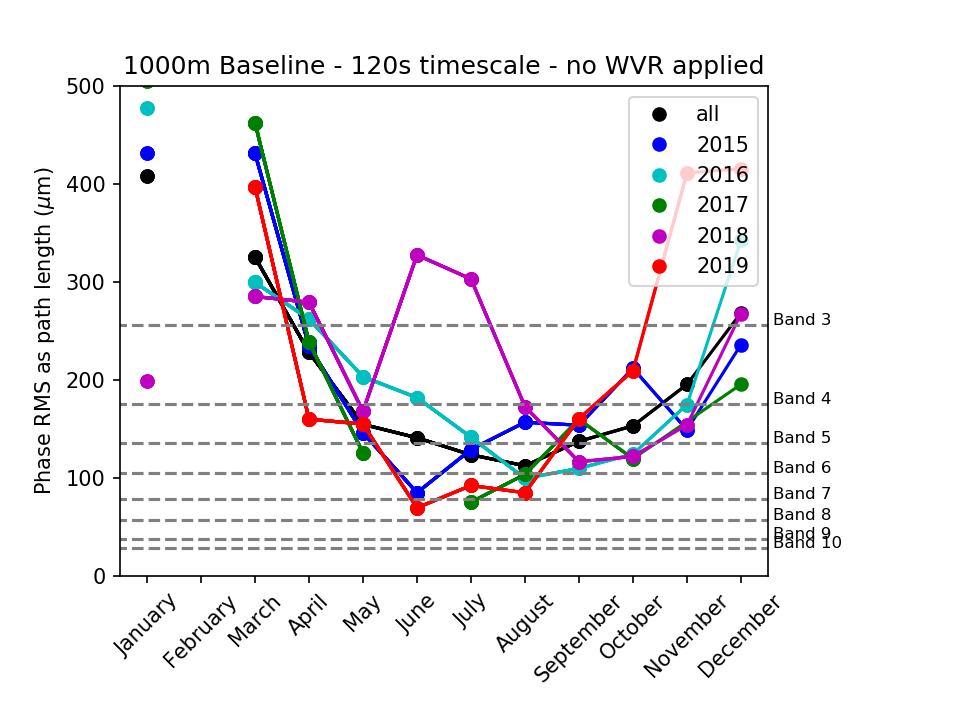}
\subcaption{Without WVR corrections}
\end{minipage}
\hfill
\begin{minipage}{0.46\textwidth}
\includegraphics[width=1.2\textwidth]{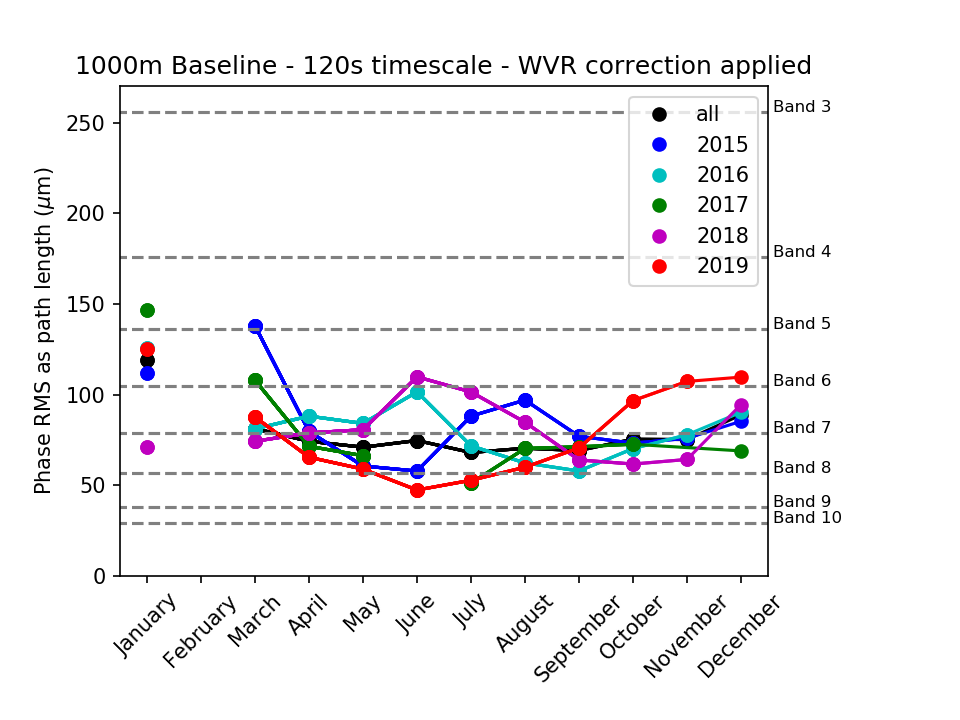}
\subcaption{With WVR corrections applied}
\end{minipage}
\caption{Variation of the median phase RMS without (left) and with WVR (right) correction as a function of month for years 2015 to 2019 inclusive. February has no data, nor June 2017. The median phase RMS can be scaled to longer or shorter baselines using the related power function from Section \ref{sec:comparable}, i.e. 1000m to 10000m with WVR applied scales the phase RMS by (10000\,m/1000\,m)$^{0.29}$ = 1.95. Note the change of y-axis scale between the panels.}
\label{fig:two1000m}
\end{figure}


In Figure \ref{fig:two1000m} we show the median phase RMS values per month for all years 2015 to 2019 inclusive without and with WVR corrections applied for the summary 1000\,m baseline length and a 120\,s timescale. As noted above, there is typically a factor of 2 change between the best and worst months (January excluded) when WVR corrections are applied but over a factor of 4 without WVR correction, although these median phase RMS values per month average out any diurnal changes (see below). The median phase RMS values shown can be scaled to longer (or shorter) baselines using the power scales indicated in Section \ref{sec:comparable}, which would result in a scaling of $\sim$1.95 from the 1000\,m to the 10000\,m baseline using the WVR corrected data.  


\begin{figure}[!ht]
\centering
\includegraphics[width=1.0\textwidth]{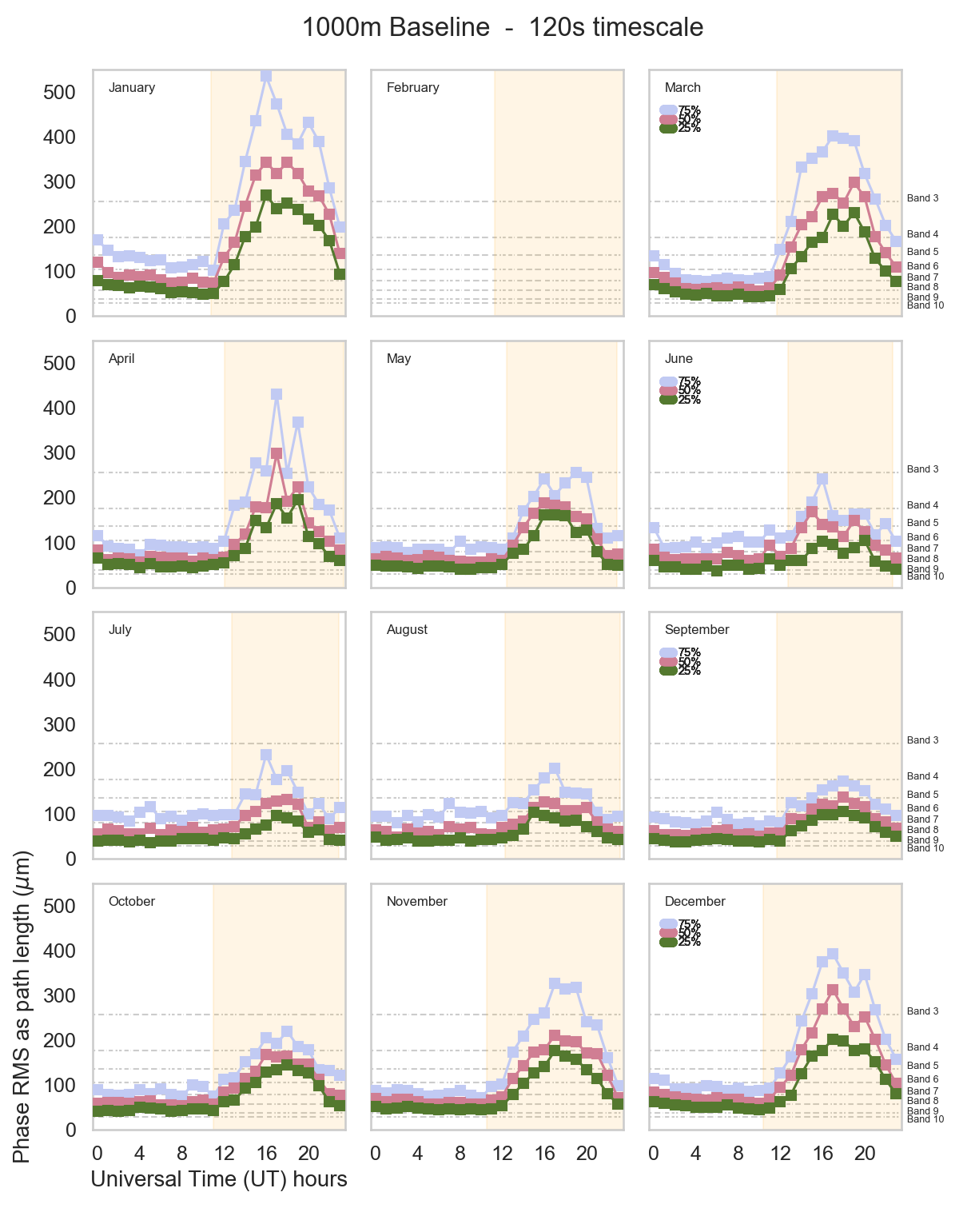}
\caption{25, 50, 75 percentiles of phase RMS, as path length variations in $\mu$m, for the summary 1000\,m baseline and 120\,s timescale after WVR corrections. The yellow bar indicates the range of sun-up and sun-down for each month. There are clear diurnal trends, with the summer showing more significant changes compared to the winter months. Like Figure \ref{fig:two1000m}, the phase RMS can be scaled to longer or shorter baselines using the related power function from Section \ref{sec:comparable}. The horizontal grey lines are the 30$^{\circ}$ limits for each band. There are no observations February.} 
\label{fig:phaseDurn}
\end{figure}

\subsubsection{Diurnal Changes}
Figure \ref{fig:phaseDurn} shows the phase RMS at the 25, 50 and 75 percentiles for data from all years binned into hourly groups for all months using the summary 1000\,m baseline and 120\,s timescale. The diurnal variation are much more significant for the phase RMS as compared to those in PWV (Figure \ref{fig:pwv24}). 
In the summer months the variation in phase RMS can change significantly, by over a factor of 7, from the night time minima to the day time peak, whereas during winter the change is less extreme with factors typically below 3. The day time in winter months typically has a similar phase RMS as compared with the night time in summer months. It is evident that the Band 9 phase RMS limit is at the 25 percentile during the night time for winter months.




\begin{figure}[!ht]
\centering
\includegraphics[width=1.0\textwidth]{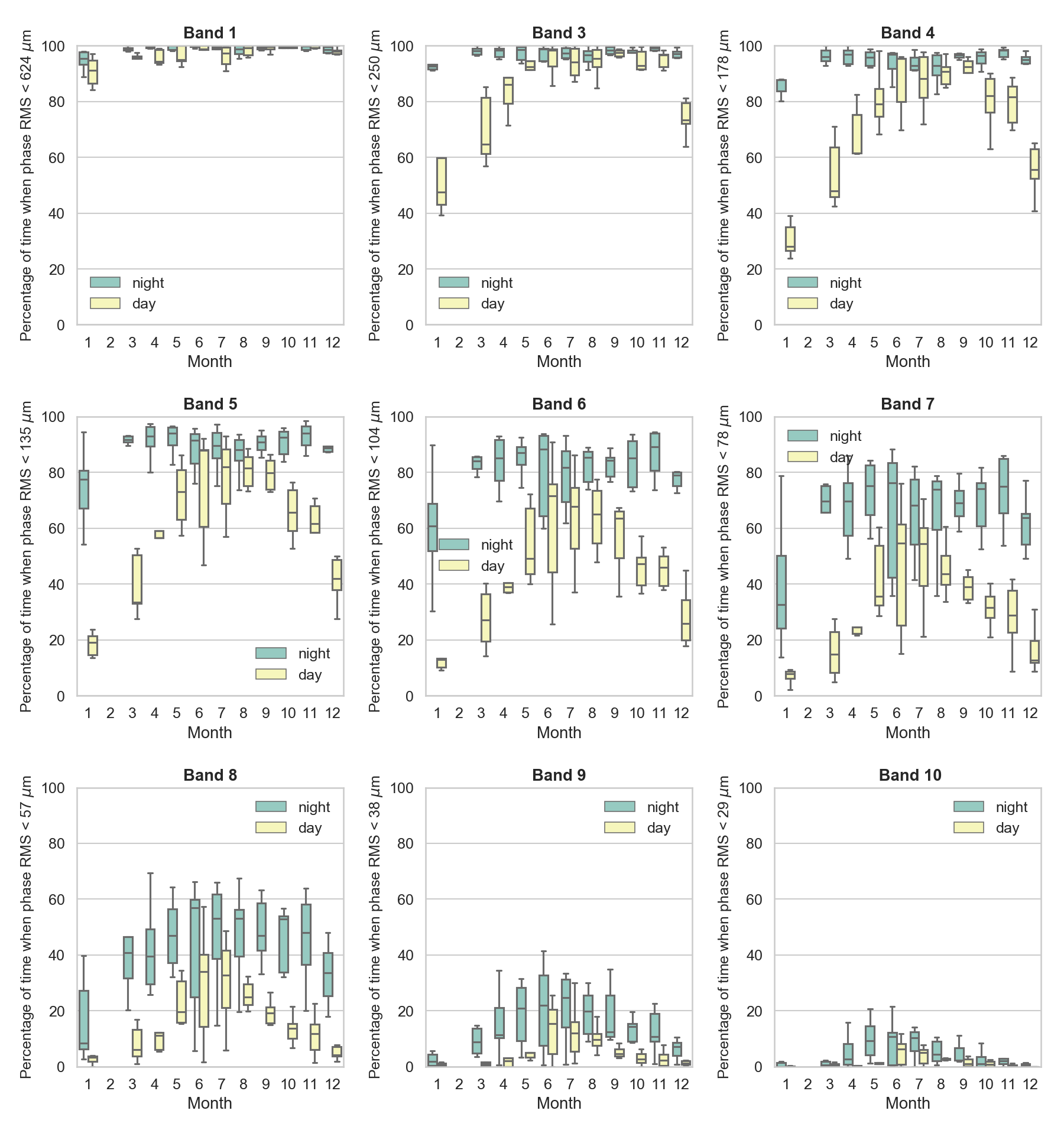}
\caption{The percentage of time when the phase RMS variability is $<$30$^{\circ}$ (shown as a path length variation in $\mu$m) at the various bands for the summary 1000m baseline and as measured over a 120s timescale. The y-axis indicates the phase RMS as a path length variation in microns. Between 1100 and 2300 UTC, equivalent to 7\,am to 7\,pm local Chilean time, is considered to be "Day" time.
The horizontal line within the box indicates to median, while the box-spread are the 25th and 75th percentiles. The whiskers show the minimum and maximal values for the best and worst year. Note, a similar plot is presented in the ALMA Proposer's Guide.} 
\label{fig:PGphase}
\end{figure}

\subsubsection{Implications for observations}

Figure \ref{fig:PGphase} is a plot following the style of the ALMA Proposer's Guide Figure 3\footnote{The Cycle 10 ALMA Proposer's Guide also includes a similar version of a phase RMS plot but where some bands are grouped together.} (also see Figure \ref{fig:fig3rep}) showing the percentage of time available for a given observing condition, however crucially here the observing conditions are based on the phase RMS levels (as path length variations) using the summary 1000\,m baseline and 120\,s timescale values. The limits are equivalent to $<$30$^{\circ}$ phase RMS for each band. Note that, unlike Figure \ref{fig:fig3rep} that shows PWV (i.e. transmission), where Band 5 (183\,GHz) is associated with Bands 9 and 10, here, in setting the phase RMS limits only the frequency of the observations matters (Equation \ref{eqn1}), and hence Band 5 is presented separately\footnote{Atmospheric dispersion can affect the path length within given bands \citep{Curtis2009}, but we do not need to consider it in calculating the representative path length variability limits per band.}. 

\begin{table}[!h]
\caption{Median percentage (to nearest percent) of time available (based on ALMA time on-sky) for each month, during day/night based on the phase RMS, as a path length variation in $\mu$m, on the summary 1000\,m baseline as measured over 120\,s. The path length cuts for the bands converts to $<$30$^{\circ}$ phase RMS per band} 

\label{tab:phpercent}
\centering
\footnotesize
\begin{tabular}{l l l l l l l l l l l l l}
\headrow \thead{Band} & \thead{Path Len.} & \thead{Jan.} & \thead{Mar.} & \thead{Apr.} & \thead{May} & \thead{June} &  \thead{July} &  \thead{Aug.} & \thead{Sep.} & \thead{Oct.} &  \thead{Nov.} & \thead{Dec.} \\    
\headrow \thead{} & \thead{($\mu$m)} & \thead{} & \thead{} & \thead{} & \thead{} & \thead{} &  \thead{} &  \thead{} & \thead{} & \thead{} &  \thead{} & \thead{} \\    
1 & 624& 91/95 & 96/99 & 94/99 & 95/100 & 99/100 & 97/99 & 99/99 & 100/99 & 100/100 & 100/100 & 98/99  \\

3 & 250& 49/93 & 69/98 & 88/98 & 92/99 & 98/99 & 94/97 & 95/97 & 98/98 & 93/98 & 97/99 & 75/97  \\

4 & 178 & 27/87 & 47/96 & 75/96 & 78/96 & 95/97 & 88/93 & 91/93 & 92/96 & 82/97 & 81/98 & 55/95 \\

5 &135 & 19/77 & 33/92 & 57/94 & 73/94 & 88/92 & 83/90 & 82/88 & 81/91 & 66/93 & 62/94 & 42/89 \\

6 & 104 & 13/61 & 27/84 & 40/85 & 51/87 & 72/88 & 68/83 & 66/86 & 64/84 & 47/85 & 49/89 & 27/79 \\

7 & 78  & 8/34 & 15/71 & 25/70 & 36/75 & 55/77 & 55/70 & 45/75 & 40/70 & 32/75 & 29/75 & 13/64 \\

8 & 57 & 3/8 & 6/41 & 11/40 & 20/47 & 34/57 & 33/53 & 25/53 & 19/47 & 14/53 & 12/48 & 4/34 \\

9 & 38  & 1/2 & 1/9 & 2/11&  5/21& 15/22&  12/25&  10/20 & 5/12  &3/14 &2/11& 1/7 \\

10 & 29  & 0/0 & 1/1 & 0/3&  1/9& 6/11&  5/10&  3/4 & 1/2  &1/1 &0/2& 0/0 \\

\end{tabular}

\end{table}
For all bands below and including Band 7 the night time indicates a flat availability of time as a function of month (excluding January), while the day time shows a notable decrease in the summer months, December thorough to March inclusive. For Bands 8, 9 and 10 both the day and night indicate a peak in time available for the winter months. The median percentages of time available are listed in Table \ref{tab:phpercent}. Band 1 observation can essentially be conducted all the time (except January), while Band 3 observations can typically be conducted $>$90\,\% of the time. At the highest frequency bands Band 9 and 10, at best, can be conducted $\sim$20-30\,\% and $\sim$10-15\,\% of the time, respectively, only in the winter months during the night time when considering the 30$^{\circ}$ phase RMS limit. At Band 10 a 40$^{\circ}$ phase RMS limit corresponds to $\sim$38\,$\mu$m, and hence the time available would be exactly as shown for Band 9.


For baselines $<$1000\,m there is a few percent increase in the time available when decreasing to the shorter, 60\,s timescale, while on the contrary there is a few percent decrease in time available moving to the longest 240\,s timescale. With respect to baseline length, for the longest summary baseline, 10000\,m, the time available reduces by $\sim$10-20\,\% for Bands 1 to 5 and by $\sim$20-30\,\% for Bands 6 and 7. At Band 8 the time available is reduced down to a maximum of 10\,\% at night in the winter months, while time available at Band 9 and 10 is effectively 0\,\%. As ALMA changes the observing scenario for different array configuration and frequencies, we discuss the changing timescales for the longer baseline lengths in Section \ref{sec:ALMAobs}. Overall, the time available for any given Band can be limited by \textit{either} the PWV or by the phase RMS, and which of these dominates depends on the ALMA array configuration. The time available based on phase RMS for the 1000\,m baseline length and 120\,s (Figure \ref{fig:PGphase}) or 180\,s timescales are well balanced with that available due to the PWV conditions. However, it is also not necessarily true that both the PWV and phase RMS criteria are met at the same time. 

\subsection{Correlations and Trends}
We highlight here that the metadata used to compare with the phase RMS are averages or standard deviations of the parameters from the entire duration of the science observations, not just over the length of the Bandpass source scan.
\subsubsection{PWV and Phase RMS}
\label{sec:PWVph}
\begin{figure}[!h]
\begin{minipage}{0.47\textwidth}
\includegraphics[width=1.03\textwidth]{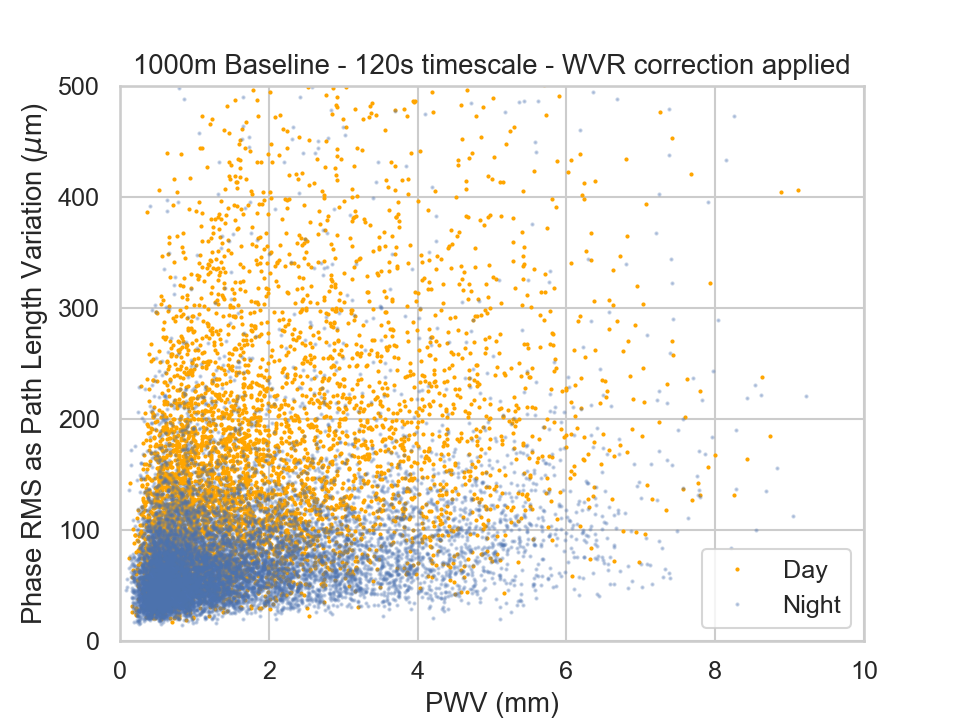}
\subcaption{PWV to 10\,mm}
\end{minipage}
\hfill
\begin{minipage}{0.47\textwidth}
\includegraphics[width=1.03\textwidth]{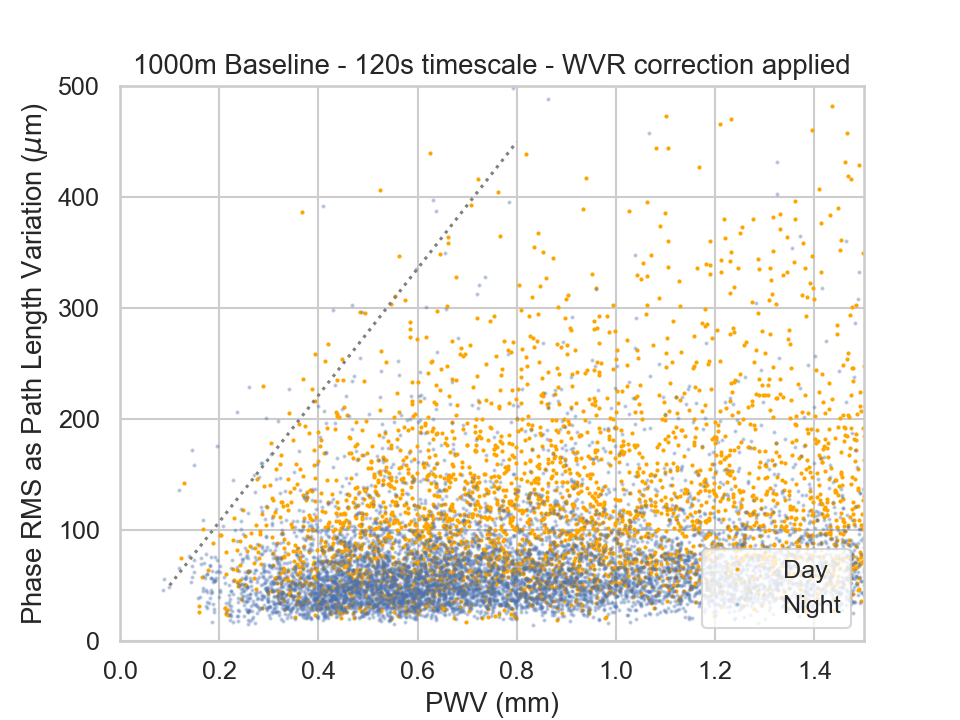}
\subcaption{PWV to 1.5\,mm}
\end{minipage}

\caption{Plots of PWV against the phase RMS (as a path length variation) for the summary 1000\,m baseline and 120\,s timescale. The right panel indicates the zoom in to low ($<$1.5\,mm) conditions. The day and night are plotted in orange and blue respectively. By-eye it is clear that night time provides lower phase RMS conditions, but that is not necessarily tied with low PWV. In panel (b) a line is drawn that could be considered to delineate the PWV - phase RMS limit, above which there are few `higher' phase RMS observations.}
\label{fig:pwvphase}
\end{figure}

Figure \ref{fig:pwvphase} shows the plot of PWV against phase RMS on the 1000\,m summary baseline length and with a 120\,s timescale (after WVR correction): the left panel indicates PWV values $<$10\,mm while the right panel zooms in to data with $<$1.5\,mm PWV. Overall there is considerable scatter but also an apparent concentration of observations towards lower PWV, $<$2\,mm, and lower phase RMS, $<$150\,$\mu$m. That said, low phase RMS can occur over a wide range of PWV values: for the 1000\,m baseline and 120\,s timescale phase RMS values $<$50\,$\mu$m were achieved in conditions with up to $\sim$5\,mm PWV, although at the very highest PWV values, $>$5\,mm, the phase RMS does not typically go below this. On the other hand, high phase RMS can be seen in observations with $<$1\,mm PWV conditions (Figure \ref{fig:pwvphase}). As has already been shown throughout this memo, the night time phase RMS is notably lower, although it is not necessarily due to the PWV being lower.

\smallskip

\begin{table}[!ht]
\caption{The range of Spearman's-rank-correlation coefficient for weather metadata, including PWV and $\Delta$PWV against the phase RMS before and after WVR correction, including all summary baselines and 60, 120, 180 and 240\,s timescales. Note the p-values are all $\sim$0, i.e. any relationships are not by chance.}
\label{tab:spearpwv}
\centering
\begin{tabular}{l l l l }
\headrow \thead{Parameter} & \thead{All} & \thead{Day} & \thead{Night} \\ 
PWV - Phase RMS (without WVR) & 0.57-0.61 & 0.50-0.54  & 0.63-0.66 \\
$\Delta$PWV - Phase RMS (without WVR) & 0.62-0.67 & 0.60-0.65 & 0.59-0.65\\
Temperature - Phase RMS (without WVR) & 0.33-0.41 & 0.34-0.41 & 0.13-0.20 \\
Wind-speed - Phase RMS (without WVR) & 0.30-0.38 & 0.27-0.33 & 0.02-0.11 \\
$\Delta$Wind-speed - Phase RMS (without WVR) & 0.19-0.25 & 0.15-0.21 & 0.06-0.12\\ 
Humidity - Phase RMS (without WVR) &0.35-0.39 & 0.38-0.42 & 0.57-0.60 \\
\hline
PWV - Phase RMS (with WVR) & 0.40-0.45 & 0.45-0.50 & 0.36-0.42 \\
$\Delta$PWV - Phase RMS (with WVR) & 0.45-0.48 & 0.50-0.53 & 0.34-0.37 \\
Temperature - Phase RMS (with WVR) &  0.23-0.29 & 0.29-0.36 & (-)0.12-(-)0.06 \\
Wind-speed - Phase RMS (with WVR) & 0.41-0.48 & 0.32-0.37 & 0.16-0.24 \\
$\Delta$Wind-speed - Phase RMS (with WVR) &  0.20-0.29 & 0.17-0.26 & 0.06-0.15\\ 
Humidity - Phase RMS (with WVR) &  0.13-0.22 & 0.23-0.32 & 0.30-0.36 \\
\end{tabular}

\end{table}

Table \ref{tab:spearpwv} indicates the ranges of the correlation coefficient parameter from a Spearman's-rank-correlation test for various weather parameters, including PWV and $\Delta$PWV (shown in Figure \ref{fig:metaphase}), with the phase RMS data with and without WVR corrections applied, for day and night time separately and combined. The PWV based correlations with phase RMS can be considered as moderate. 
The coefficients indicate that $\Delta$PWV correlations could be marginally stronger, except at night time (particularly after WVR correction). This possibly points to different atmospheric behaviour, such as solar heating creating more turbulence during the day that is tied with a higher variability in PWV, although a deeper investigation on the weather conditions is beyond the scope of this memo. 
\citet{Evans2003} reported a `significant' correlation between PWV and phase RMS despite a large scatter in the data. They noted that when the transparency is better than the median the phase RMS was `twice as good as otherwise'. This roughly ties with the phase RMS data without WVR corrections in our sample. The median phase RMS is $\sim$200\,$\mu$m (1000\,m baseline, 120\,s timescale), whereas the median phase RMS of data below the median PWV (1.24\,mm) is $\sim$115\,$\mu$m, i.e. almost twice as good. The same holds for day and night independently, where the PWV medians are $\sim$1.4\,mm and $\sim$1.1\,mm respectively, and the phase RMS ratios change by $\sim$1.6. Note that irrespective of the phase RMS absolute value, dependent on the baseline length and timescale, the trends are still the same as shorter baselines and timescales simply have lower absolute phase RMS, while longer baselines and timescales have higher absolute phase RMS values. 

Overall, lower PWV does tend to provide a higher concentration of lower phase RMS observations, however, it is still possible to have poor phase RMS while the transmission is good. In Figure \ref{fig:pwvphase}, more clearly in panel (b) - dotted line, there is a possible upper limit to the phase RMS as a function of PWV as can be delineated by a diagonal drawn from $\sim$0.1\,mm PWV and 50\,$\mu$m phase RMS up to $\sim$0.8\,mm PWV and 450\,$\mu$m phase RMS. Above the line there are very few high phase RMS observations. The limit does not help particularly with ALMA high frequency observations where $<$79\,$\mu$m is required for Band 7 or higher because values above this can occur when PWV $>$0.2\,mm, well below the nominal Band 10 PWV value (Table \ref{tab:octile}). For longer baselines the phase RMS of the observations, and the dotted line move to larger values.

\subsubsection{Weather parameters with Phase RMS}
\label{sec:metaph}

\begin{figure}[!ht]

\begin{minipage}{0.47\textwidth}
\includegraphics[width=1.0\textwidth]{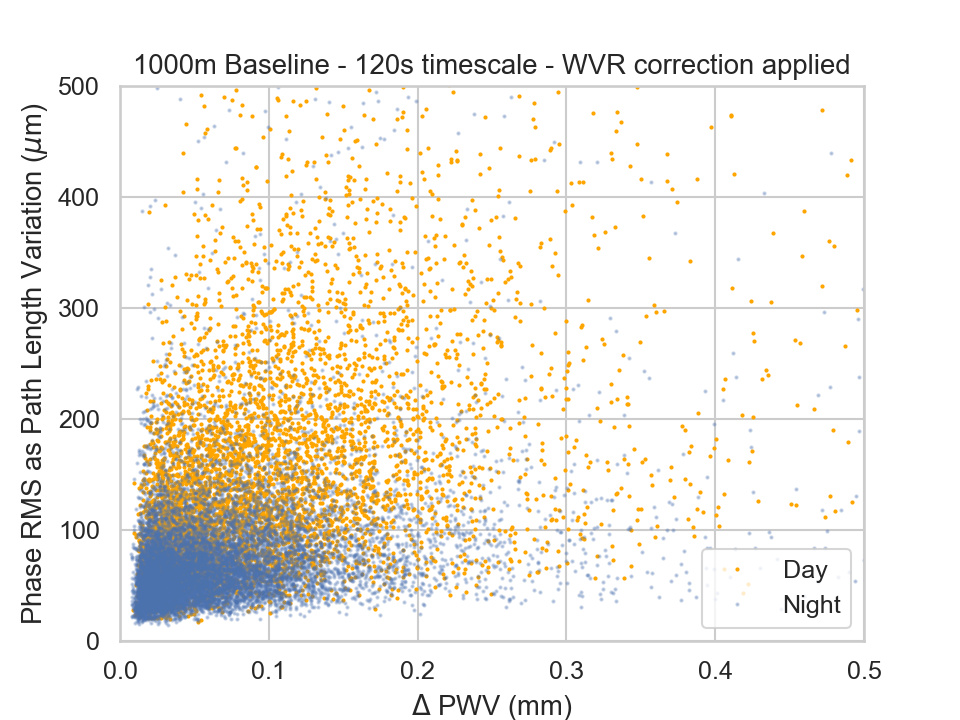}
\end{minipage}
\hfill
\begin{minipage}{0.47\textwidth}
\includegraphics[width=1.0\textwidth]{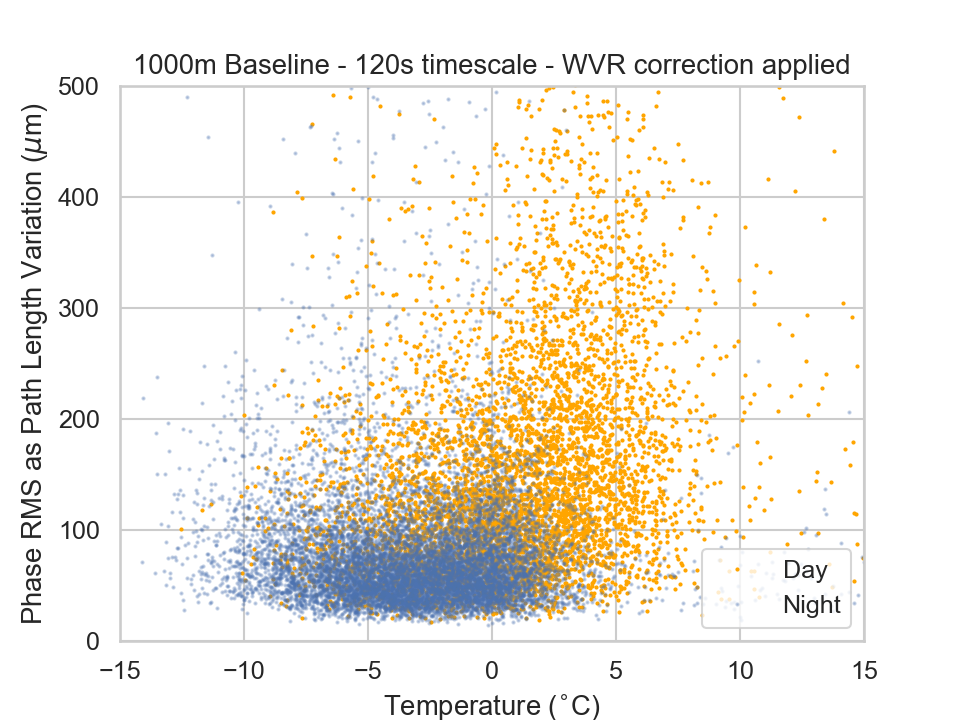}
\end{minipage}
\hfill
\begin{minipage}{0.47\textwidth}
\includegraphics[width=1.0\textwidth]{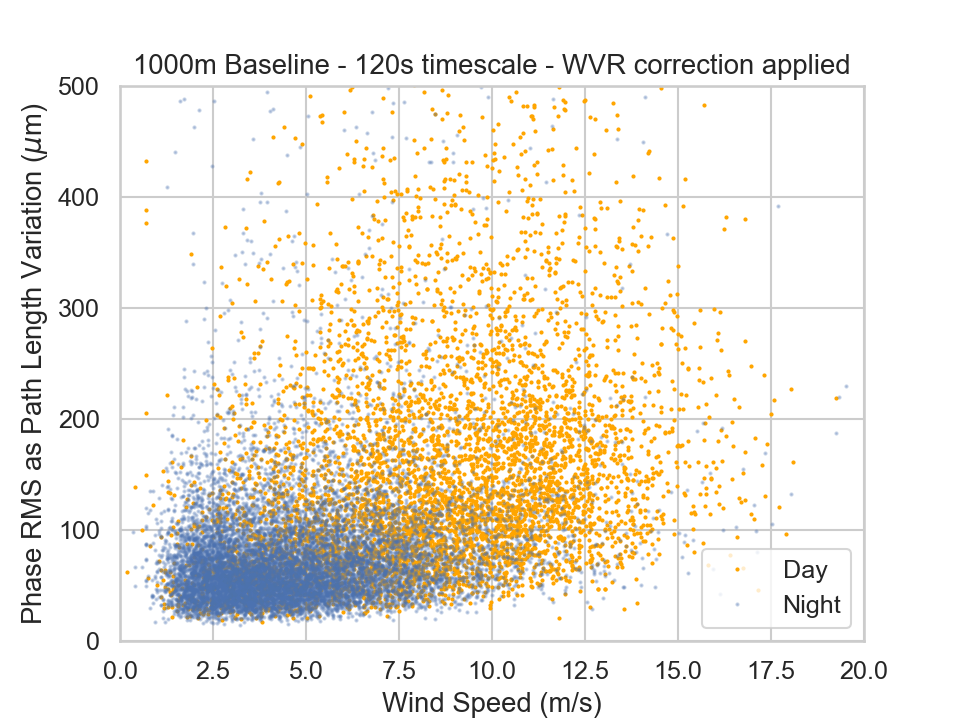}
\end{minipage}
\hfill
\begin{minipage}{0.47\textwidth}
\includegraphics[width=1.0\textwidth]{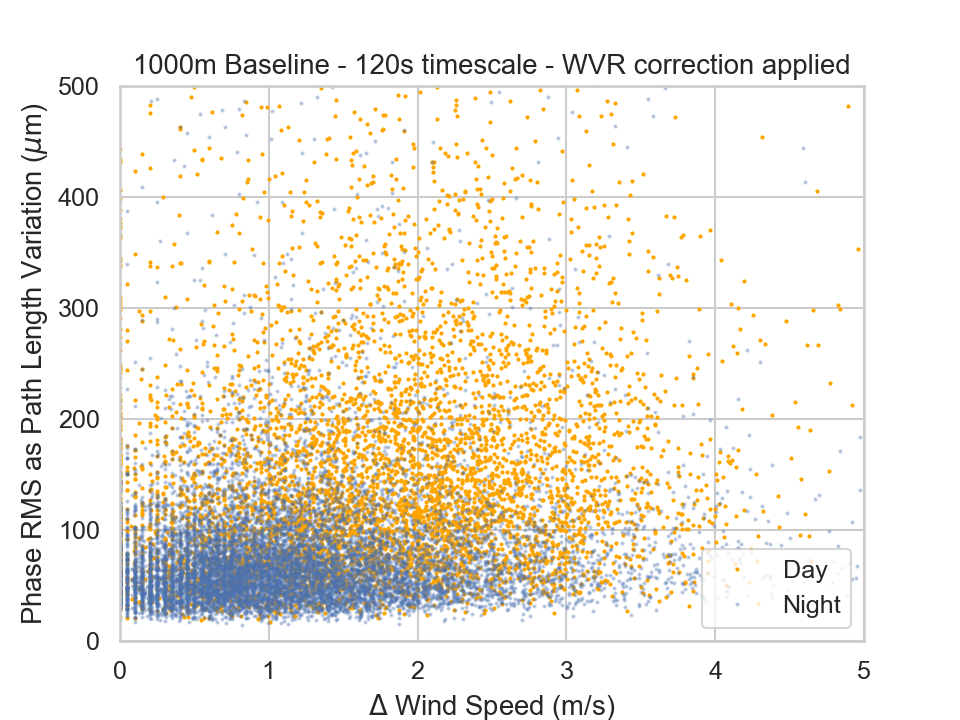}
\end{minipage}
\hfill
\begin{minipage}{0.47\textwidth}
\includegraphics[width=1.0\textwidth]{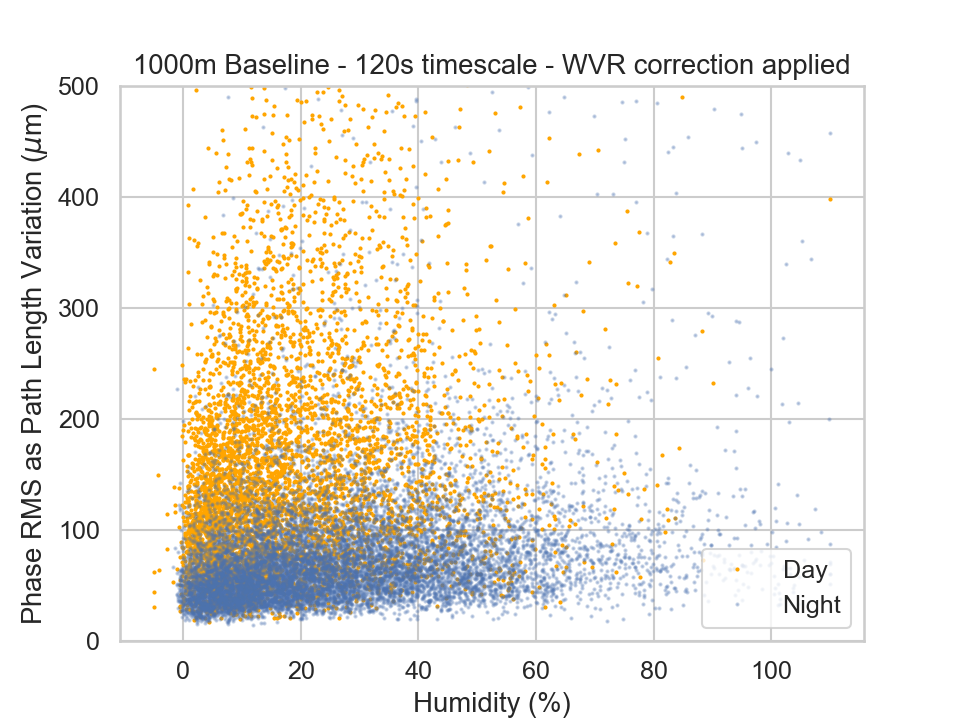}
\end{minipage}
\hfill
\begin{minipage}{0.47\textwidth}
\includegraphics[width=1.0\textwidth]{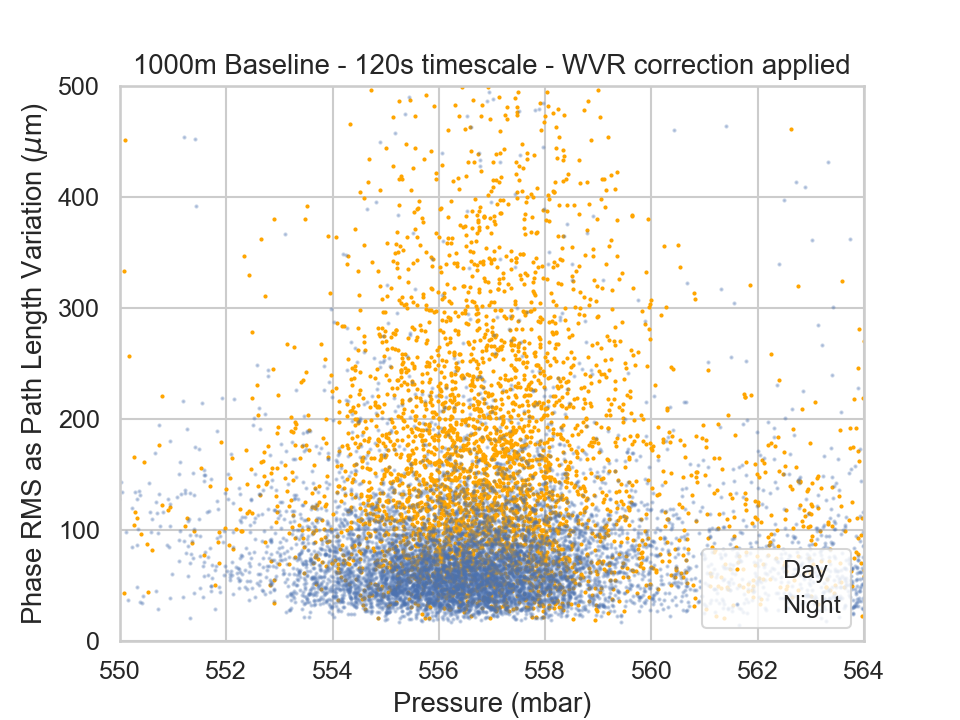}
\end{minipage}

\caption{The six panels show the $\Delta$PWV, temperature, wind speed, $\Delta$wind speed, humidity and pressure against the phase RMS, as a path length variation, for the summary 1000\,m baseline as measured over 120\,s with WVR corrections applied. Both lower temperatures and lower wind speeds tie with lower phase RMS values and predominantly occur at night time.}
\label{fig:metaphase}  
\end{figure}

In the six panels in Figure \ref{fig:metaphase} we plot the $\Delta$PWV, temperature, wind speed, $\Delta$wind speed, humidity and pressure metadata against the phase RMS, again on the 1000\,m baseline measured over a 120\,s timescale. Here we echo the weather trends seen by \citet{Evans2003}: lower phase RMS (with and without WVR correction) is concentrated towards lower temperatures, of which occur primarily at night and at lower wind speeds. 

Making cuts to separate the observations with temperatures above and below the median temperature of $-$1.527$^{\circ}$C, we find that there are approximately 1.5 times more observations below the limit. This is closely tied with the ratio of day to night observations, although the cut does not separate day and night exclusively as $\sim$20\,\% of the observations with temperatures $<-$1.527$^{\circ}$C were made in the day time (1100-2300 UTC). The corresponding median phase RMS values (with WVR correction applied) for observations below and above the median are $\sim$70\,$\mu$m and $\sim$110\,$\mu$m respectively. 
Similarly, in making cuts for wind speed below and above 10\,m/s we find the median phase RMS (after WVR correction) values are $\sim$70\,$\mu$m and $\sim$140\,$\mu$m respectively, pointing to high winds linking with a more turbulent atmosphere. Lower winds speeds also predominately occur at night, where there is obviously no solar heating, although $\sim$85\,\% of the observations have a wind speed $<$10\,m/s.

Humidity also appears to have a larger concentration of low phase RMS observations at lower humidity, $<$20-30\,\%. A stronger correlation is also reported at night time as compared to the day. In examining the median values of the phase RMS at night time, 67, 76, 79, 85, 85\,$\mu$m for each 10\,\% bin up to 50\,\%, there is only a slight increase with humidity. 



Based on the Spearman's-rank-correlation coefficients reported in Table \ref{tab:spearpwv} the correlations range from none to weak/moderate, and slightly differ when considering all the observations or sub-sets of day and night. We summarize these below:
\begin{itemize}
    \item  Temperature is weakly/moderately correlated with phase RMS for the full sample and day time, likely due to solar heating and an increase in turbulence. There is no significant trend at night time which is predominately limited to lower temperatures all of which have lower phase RMS. 
    \item Wind is weakly/moderately correlated with phase RMS. The correlation is stronger after WVR corrections are applied. Observations with lower wind speeds, $<$10\,m/s, do indicate a factor of two reduction in the median phase RMS. Similar to temperature, night time observations consists of a sub-sample of lower wind speeds that have lower phase RMS values - therefore weakening the night time correlation. There is no trend with wind direction (not shown). 
    \item Humidity shows a weak correlation with phase RMS after WVR application. In comparing the median phase RMS values there is a slight trend of increasing values with increasing (binned) humidity, however the magnitude of the change in median phase RMS is largely outweighed by the general spread of phase RMS.
    \item There are no correlations with atmospheric pressure. The largest correlation coefficient is 0.13 (not tabulated).
\end{itemize}
\smallskip

\noindent Ultimately, irrespective of the correlations we cannot directly use the relationships to estimate the phase RMS of a given observation based solely on the metadata.

\subsubsection{Other Weather Correlation}
\label{sec:metawe}
For completeness we briefly compare the weather metadata as a whole with all other metadata values. As reported by \citet{Evans2003}, we find westerly winds are stronger than the easterly ones, but note that the wind is usually blowing westerly for the majority of the observations. Furthermore, the humidity and PWV parameters have an apparent correlation, naturally more humid conditions point to a higher likelihood of high water vapour content. Wind speed and $\Delta$wind speed are correlated, in that higher wind speeds have the potential to be the most variable. 


\subsection{WVR correction}
\label{sec:wvrcorr}
A significant update in this memo compared with \citet{Evans2003} is that the water vapour radiometer (WVR) system\footnote{Also see ALMA memos 252 and 361.}  \citep{Hills2000,Hills2001,Hills2004,Nikolic2013} is no longer theory, but has been a working system for over 10 years at ALMA. As already noted earlier, the phase RMS values are significantly improved after application and therefore it is assumed that the WVR corrections will be applied during calibration\footnote{The ALMA calibration pipeline investigates whether the phase RMS is actually reduced when the WVR corrections are applied. Typically there is always an improvement, but in rare cases where the phase RMS does not improve the Pipeline does not apply the WVR solutions.}.

\begin{figure}[!ht]
\begin{minipage}{0.47\textwidth}
\includegraphics[width=1.05\textwidth]{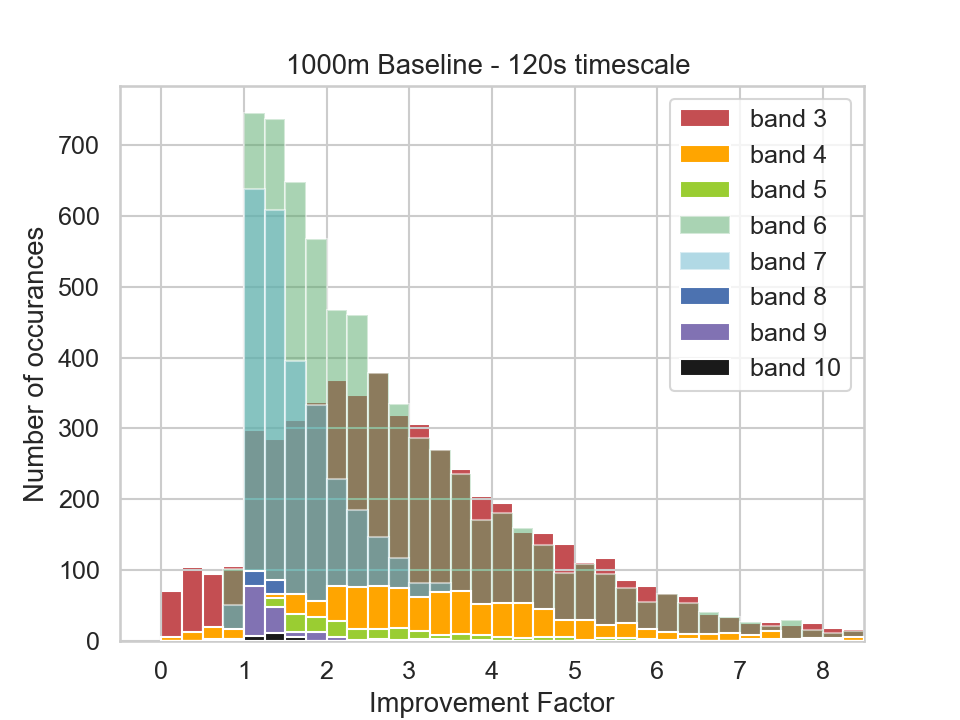}
\end{minipage}
\hfill
\begin{minipage}{0.47\textwidth}
\includegraphics[width=1.05\textwidth]{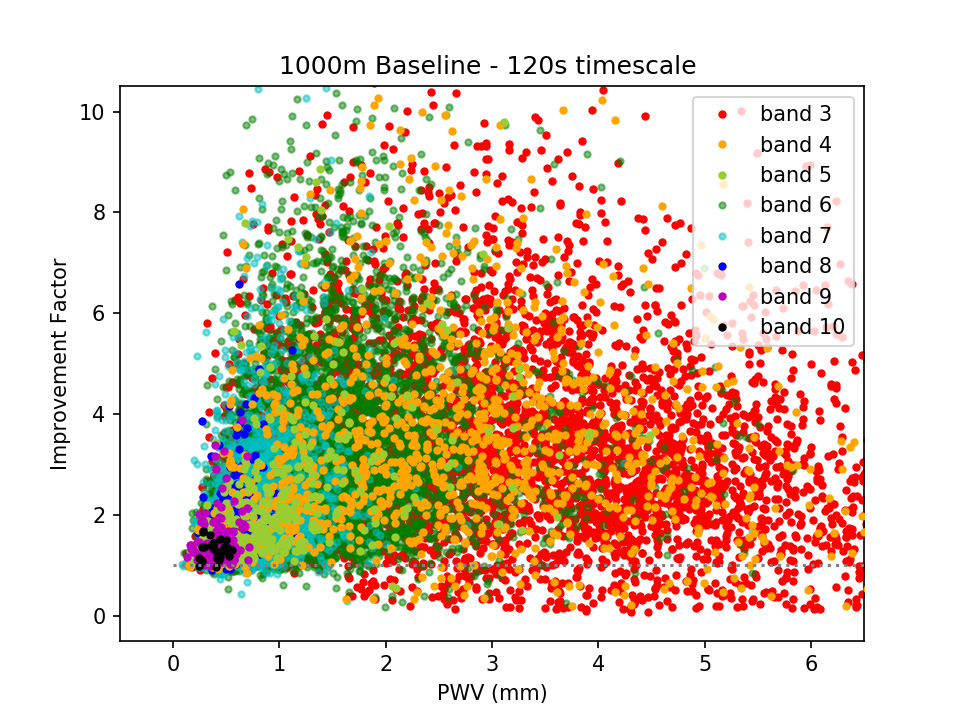}
\end{minipage}

\caption{Left - Improvement factor for the phase RMS after applying the WVR corrections. Right - Improvement factor as a function of PWV. Colour indicates the observing band used. Note that the brown colour is due to the overlap of Bands 3, 6 and 7. The vast majority of data ($>$96\,\%) have some improvement after WVR application ($>$1), while those without improvement are mostly related to Band 3. Generally, lower frequency bands, observed in higher PWV conditions have a larger median improvement factor, although there is considerable scatter in improvement factor against PWV, and some low PWV observations can also exhibit a large correction, $>$2-3.}
\label{fig:wvrhist}
\end{figure}

The left panel of Figure \ref{fig:wvrhist} shows the histogram of the \textit{Improvement Factor} of the phase RMS after application of the WVR correction when considering the phase RMS from the 1000\,m summary baseline length and as measured over 120\,s. The symbols are coloured as a function of the observing band. The distribution of the improvement factors remain largely unchanged irrespective of the summary baseline length or timescale used. Firstly, the vast majority of the observations ($>$96\,\%) are improved by the WVR application and have improvement factors $>$1; secondly, over half of the observations ($>$58\,\%) show a phase RMS improvement of a factor of 2 or more; thirdly, the majority of observations that did \textit{not} improve after WVR application were at low frequency bands, mainly Band 3, but also for those at Band 6. Within the nearest percent there is no change in the number of observations improved by WVR application for changing timescales, but there is a decrease to 92\,\% and 90\,\% of observations with an improvement factor $>$1 for 5000\,m and 10000\,m summary baseline lengths, respectively. 

The right panel of Figure \ref{fig:wvrhist} shows the spread of improvement factors as a function of PWV. For observations taken using Band 3 and 4 the improvement factor median values are between 2.5 to 3.0 and 2.6 to 3.2 considering all summary baselines from 500\,m to 10000\,m and 60\,s to 240\,s timescales, respectively, while observations in all bands higher than Band 6 have median improvement ratios $<$2.0 (see Table \ref{tab:wvrrat}). Of course lower frequency bands can be observed in higher PWV conditions, whereas higher frequency bands are necessarily restricted to lower PWV. Observations in the lowest PWV conditions do tend to have a lower improvement factor after WVR application, although there is still a notable scatter and even data taken in 0.5-1.0\,mm PWV can have improvement factors $>$2.

There are roughly 4\,\% of all observations that do not improve after WVR application (for the 1000\,m summary baseline length and 120\,s timescale), of which only $\sim$2\,\% of those used Band 8, 9 or 10\footnote{As noted in the Appendix, the database is not 100\,\% complete and there are some tens of high frequency observations that have not yet been ingested. However, we do not expect the percentages to change significantly as low observing band observations with higher PWV dominate those that do not improve after WVR application.}. The median PWV for these Band 8, 9 and 10 observations is 0.37\,mm, with a maximum of only 0.55\,mm, indicative of very good transmission conditions as one would expect for high frequency observations. The median phase RMS, as a path length variation, is $\sim$50\,$\mu$m before WVR corrections are applied (1000\,m baseline, 120\,s timescale) and point to the best stability conditions. The median improvement factor for these observations is 0.98, and hence the WVR application essentially has no effect, rather than making the phase RMS noticeably worse. 

On the contrary, the majority of observation for Band 3 to 6 that get worse, some by more than a factor of 2 after WVR application, were taken in \textit{higher} PWV conditions ($>$1.5-2.0\,mm). In these unique cases it has been found that in the presence of clouds above ALMA the `continuum' emission from liquid water (hydrosols) dominates the WVR brightness temperature, which is used for generating the phase corrections, and hence the WVR solutions are erroneous. Software has since been developed, {\sc remcloud}\footnote{See the ALMA technical handbook and the ALMA Knowledgebase Article: https://help.almascience.org/kb/articles/what-is-remcloud-and-how-could-it-reduce-phase-rms.}, that acts to remove the cloud continuum from the WVR spectral windows and allows for a more optimal functioning of the WVR system. 


\begin{table}[!t]
\caption{Range of quartiles of the WVR improvement ratio as a function of observing band considering baseline lengths between 500\,m to  10000\,m and 60\,s to 240\,s timescales. Note very high outlier phase RMS values beyond the 95$^{\rm{th}}$ percentile are excluded, while improvement factors $<$1 are still included. }
\label{tab:wvrrat}
\centering
\begin{tabular}{l l l l}
\headrow \thead{Quartile:} & \thead{25}  & \thead{50} &  \thead{75}  \\
Band 3  &  1.7-2.1 & 2.5-3.0 & 3.5-4.3 \\
Band 4 &  1.8-2.1 & 2.6-3.2 & 3.6-4.4 \\
Band 5  &  1.2-1.4  & 1.7-2.0 & 2.6-3.0 \\
Band 6  &  1.3-1.6 & 2.0-2.4 & 3.0-3.6 \\
Band 7 &  1.1-1.3 & 1.4-1.7 & 2.0-2.4 \\
Band 8 &  1.0-1.2 & 1.3-1.5 & 1.6-1.9 \\
Band 9 &  1.0-1.1 & 1.1-1.3 & 1.2-1.6 \\
Band 10 & 1.0-1.3 & 1.1-1.3 & 1.2-1.5 \\
\end{tabular}
\end{table}

\section{Observations with ALMA}

\subsection{The Importance of Low Phase RMS}
\label{sec:impphase}
The level of the phase RMS that remains in a target source after all calibration processes have been made will govern the quality and accuracy of the final data, and images. With phase fluctuations remaining in the data after calibration the coherence for interferometric visibilities (${\bf V} = {\bf V}_0e^{i\phi}$, where ${\bf V}_0$ represents the true visibilities and $\phi$ are the phase fluctuations caused by the troposphere, \citealt{Thompson2017}) can be calculated following:


\begin{equation}
  \label{eqn2}
\langle {\bf V} \rangle = \langle {\bf V}_0 \rangle \, \langle e^{i\phi} \rangle = \langle {\bf V}_0 \rangle \, e^{-\sigma^2_{\phi}/2},
\end{equation}

\noindent assuming Gaussian random phase fluctuations, $\phi$, with a \textbf{Phase RMS} of $\sigma_{\phi}$ (in radians) and a zero degree mean phase. The parameter $\sigma_{\phi}$ is the Phase RMS as we report throughout this memo as a path length variation ($\ell$, in $\mu$m). 

\begin{figure}[!ht]
\centering
\includegraphics[width=0.95\textwidth]{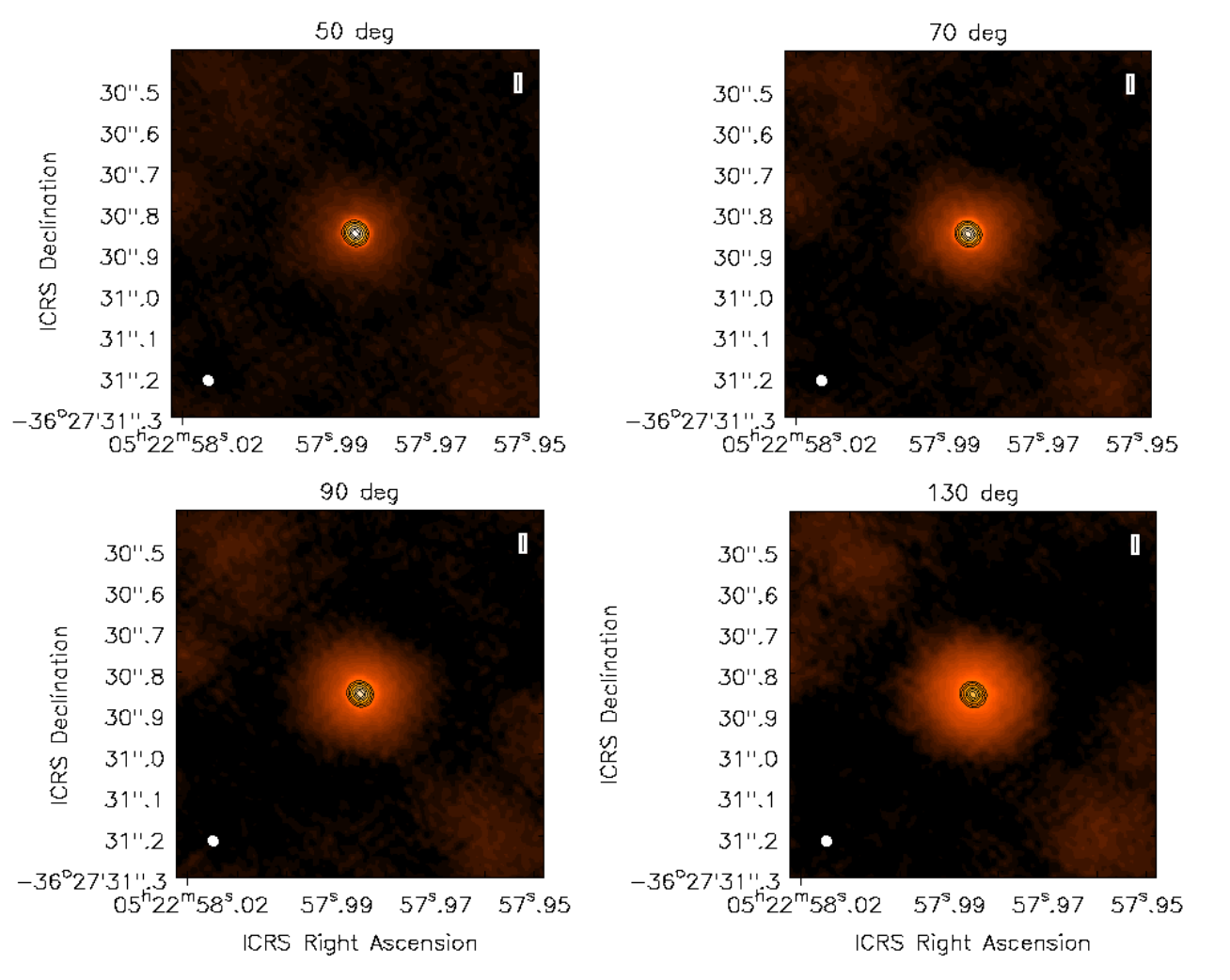}
\caption{Ideal 1\,Jy flux point source as corrupted by simplistic Gaussian phase noise peaking at 50$^{\circ}$, 70$^{\circ}$, 90$^{\circ}$ and 130$^{\circ}$ phase RMS levels on the longest baselines, left to right, top to bottom. With increasing corruptions the flux is spread further away from the source and the effective resolution of the observations becomes worse. A point source should be unresolved within the beam (bottom left), without any larger scale emission. The 50$^{\circ}$ phase RMS could be considered as an `adequate' observing condition. The black contours are the 5, 10, 20, 40, 60, 80\,\% levels of the same ideal point source corrupted with $\sim$30$^{\circ}$ phase RMS for comparison. Peak flux densities of these images are 0.83, 0.69, 0.54 and 0.29\,Jy/bm. The simulated observations use Band 6 long baselines and have a synthesised beam of $\sim$30\,mas.} 
\label{fig:fakephase}
\end{figure}

A 30$^{\circ}$ good-to-ideal phase RMS limit relates to a coherence value of 0.87, i.e. 87\,\% of the amplitude is correctly attributed to the target, the rest is decorrelated flux, effectively spread around the image due to phase errors\footnote{If targets are strong enough and relatively compact phase self-calibration can be performed that can correctly attribute the flux spread around an image to the target as further phase corrections are possible \citep{Brogan2018,Richards2022}.}. For simplicity, for a point source, the peak flux density will drop whereas the integrated flux over a region around the source, \textit{larger} than the synthesised beam size, would typically recover the majority of the total flux of the source. The spreading of flux, due to phase errors is effectively a `seeing' for sub-millimeter interferometers and thus worsening phase RMS means the expected resolution is not actually achieved. Figure \ref{fig:fakephase} shows scenarios where an ideal 1\,Jy (total flux) point source has been corrupted with simplistic Gaussian phase noise peaking at 50$^{\circ}$, 70$^{\circ}$, 90$^{\circ}$ and 130$^{\circ}$ phase RMS at the longest baselines and where phase RMS was increased as a function of baseline length (note that a phase RMS of 50$^{\circ}$ can be considered as `adequate' conditions, while $>$70$^{\circ}$ are very poor). The figure illustrates the spreading of the source flux away from a point-like structure and how the effective angular resolution is worse (larger) than the associated synthesised beam. These corrupted images are ideal cases where the emission is evenly spread due to the false Gaussian like phase noise on all baselines, whereas true phase errors can include systematic offsets and the stochastic variations caused by the atmosphere - we refer the reader to Figure 5 of \citet{Carilli1999}, and Figure 5 of \citet{Maud2022} where images of point sources are shown with worsening levels of \textit{real} phase errors. 



Recently, investigations to open the long-baseline and high-frequency observation capabilities of ALMA reported that 0.7 is a pragmatic image coherence\footnote{The peak flux density divided by the integrated flux, valid for unresolved or point-like sources, this is also referred to as the fractional peak flux recovered \citep[e.g.][]{Dodson2009,Rioja2011,Rioja2015,MartiVidal2010b}} level \citep{Asaki2020,Asaki2020b,Maud2020,Maud2022}. For their point source test targets the image defects, away from point-like structure (i.e. beyond the Gaussian shape of the synthesised beam), are negligible to minimal for images with $>$0.7 image coherence. For an image coherence below 0.5-0.6, the image defects and flux decorrelation at significant levels, $>$3-5 sigma, are apparent. Considering the phase error budget, that is the combination of atmospheric phase variations and the systematic errors imparted due to the calibrator-to-target separation angle, \citet{Maud2022} note that \textbf{the phase RMS should ideally be $<$30-40$^{\circ}$} (good atmospheric stability), along with the use of close phase calibrators (reducing systematic errors), within a few degrees, as to achieve an image coherence $>$0.7. Importantly, regardless of what array configuration is used, or the phase referencing cycle time, the phase RMS should always meet the required limit such that the target source is not excessively corrupted. As already shown in Section \ref{sec:phasemain}, this is considerably easier for lower frequencies where overall larger path length variations (Equation \ref{eqn1}) correspond to relatively small phase RMS levels.

\subsection{ALMA's current observing schemes}
\label{sec:ALMAobs}
Because of intrinsic phase variations increasing with timescale, observing frequency and to some extent also with baseline length\footnote{As noted in Section \ref{sec:comparable}, for short timescales the phase RMS with baseline length does not necessarily continue to increase moving to baselines $>$1000-5000\,m.} for any given atmospheric condition, ALMA employs different observing strategies as a function of array configuration and frequency. These are also continually being updated, most recently for Cycle 10 at higher frequency bands (8, 9 and 10) in part from the results presented in this memo (see Section \ref{sec:optim}). 

ALMA employs gradually faster cycle times, i.e. more frequent visits to the phase calibrator, as a function of increasing observing bands and for different array configuration groups. The following schemes are those used for Cycle 9 observations (since October 2022). For the shortest baseline configurations C-1 to C-4 ($<$784\,m) the target scan time is reduced down from 10\,min to 5\,min for Band 3 to Band 10. Although we do not investigate timescales $>$240\,s in this memo, as described in Section \ref{sec:overallph}, the phase RMS for short baseline arrays essentially saturates as these configurations are not sensitive to spatially larger turbulent atmospheric cells, meaning that the phase RMS will not noticeably increase even when increasing the timescales beyond 5-10\,min at the lowest observing bands. It could be possible to increase cycle times if phase stability is already good as to improve efficiency (Section \ref{sec:optim}).

For mid-baseline length configurations, C-5 (1.4\,km) and C-6 (2.4\,km), the target scan times are reduced from 5\,min down to 3\,min, again moving from Band 3 to Band 10. Note that the phase calibrator scan lengths are typically 30\,s, but can be 60\,s for higher frequencies where sources are weaker and a sufficient SNR is required for the calibrators. As an example, using the Phase RMS database we compare 180\,s (3\,min) and 240\,s (4\,min) timescales and find $\sim$5-10\,\% more time available (in winter months) for Band 9 and 10, in terms of meeting the 30$^{\circ}$ phase RMS limit (i.e. the strategy of shorter cycle times in ALMA operations is validated). If even faster cycle times were possible, there would be a further increase in time available, although this comes at the expense of higher overheads (Section \ref{sec:optim}). 

\begin{figure}[!ht]
\centering
\includegraphics[width=1.0\textwidth]{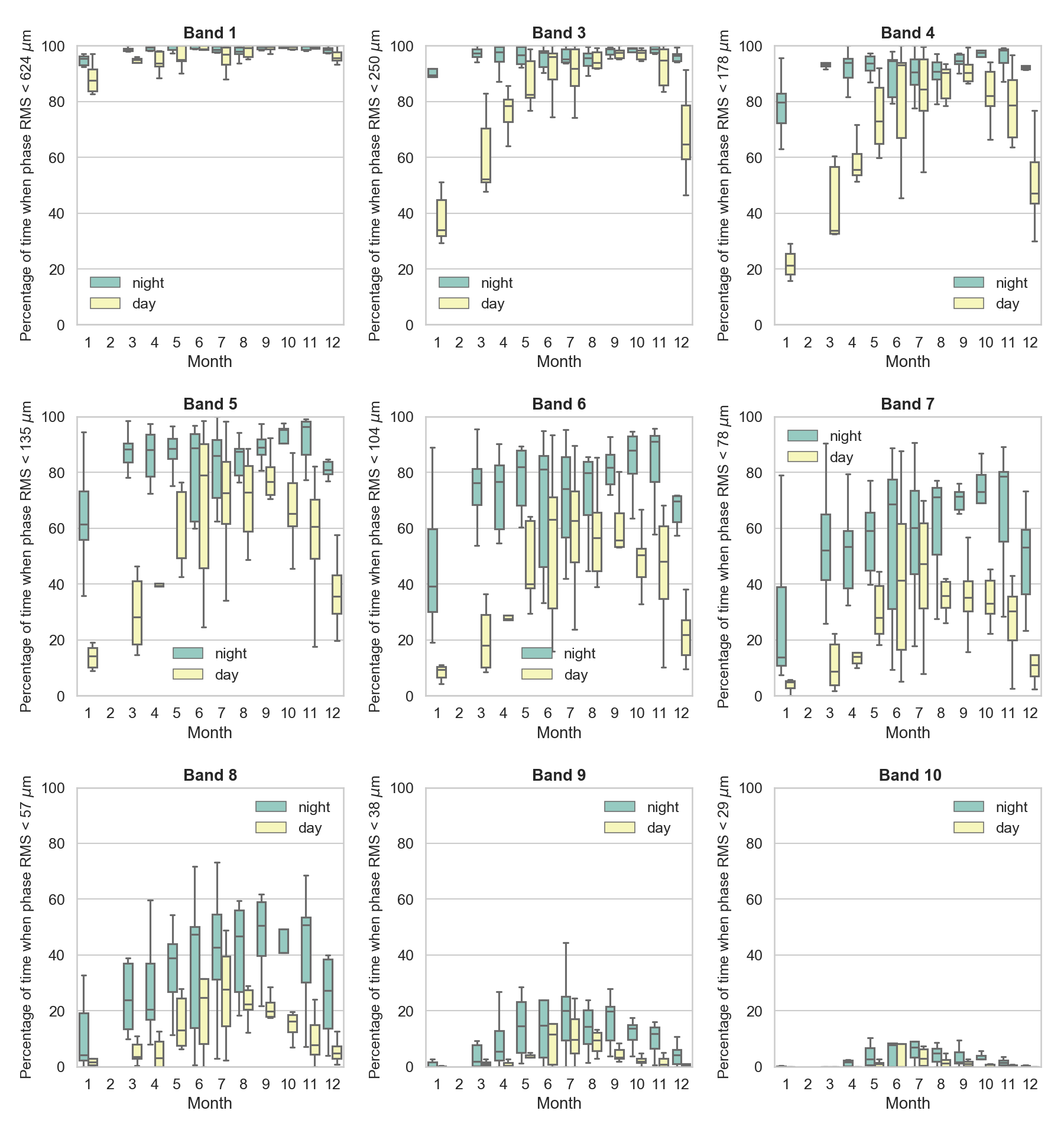}
\caption{The percentage of time when the phase RMS variability is $<$30$^{\circ}$ (shown as a path length variation in $\mu$m) at the various bands for the summary \textbf{5000m} baseline length and as measured over a \textbf{60\,s} timescale which closely ties with the C-7 ALMA configuration and the 54\,s target scan time (compare with e.g. Figure \ref{fig:PGphase}). The y-axis indicates the phase RMS as a path length variation in microns. Between 1100 and 2300 UTC is considered to be "day" time. The horizontal line within the box indicates to median, the box-spread are the 25th and 75th percentiles, and whiskers are the minimum and maximum values per month considering all years.} 
\label{fig:5000_60s}
\end{figure}

For the C-7 (3.6\,km) configuration a 54\,s target scan time is used for all bands. Observing efficiency is not reduced significantly compared to the shorter baseline configurations because the scan time of the phase calibrator is also reduced to maintain a reasonable duty cycle (18\,s is used, so the cycle times are $\sim$76\,s including 2\,s slewing\footnote{Note that `slewing' encompasses not only the antenna movement ($<$2\,s), but antenna and software stabilisation.} to and 2\,s slewing from the target). Using these short target scan times means that ALMA can observe at long baselines $>$90\,\% of the time for Band 3 to Band 5 at night in all months other than January and December, and $>$70-80\,\% of the time during the day in winter months (Figure \ref{fig:5000_60s}), based upon the 30$^{\circ}$ phase RMS limit. For Bands 8 and 9 the time available based on the phase RMS limit is reasonably coincident with that available based on meeting the nominal PWV (compare with Figure \ref{fig:fig3rep}), $\sim$40\,\% and $\sim$10-20\,\% at night time in the winter months, respectively. Band 10 observations are the most difficult to conduct given the short amount of time available that meets the 30$^{\circ}$ limit. If however a 40$^{\circ}$ limit was acceptable (acknowledging the slight drop in coherence) then the time available matches that for Band 9, that is almost doubled from $\sim$5-7\,\% to $\sim$10-15\,\% at night time for winter months (i.e. 38\,$\mu$m $\sim$40$^{\circ}$ at Band 10). Recall that in Section \ref{sec:comparable} we obtain the longer summary baseline phase RMS values by scaling from shorter summary baseline lengths for most EBs, and hence we may overestimate the phase RMS for the 60\,s timescale and therefore we provide a more pessimistic view of the time available.

\begin{figure}[!ht]
\centering
\includegraphics[width=1.0\textwidth]{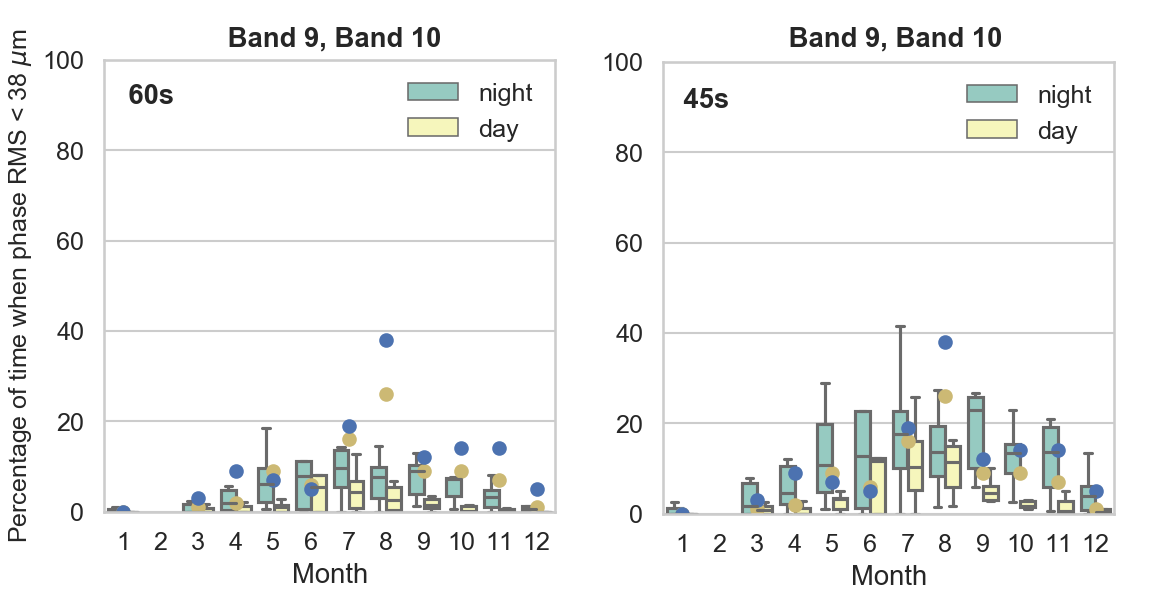}
\caption{The percentage of time when the phase RMS variability is $<$30-40$^{\circ}$ at Band 9/10 for the summary \textbf{10000m} baseline length, that closely ties with the C-8 to C-10 ALMA configurations, and as measured over a \textbf{60\,s} timescale (left) and a \textbf{45\,s} timescale (right). The y-axis indicates the phase RMS as a path length variation in microns. Between 1100 and 2300 UTC is considered to be "day" time. The horizontal line within the box indicates to median, while the box-spread are the 25th and 75th percentiles and whiskers the minimum and maximum values per month considering all years. The circles represent the median time available based on meeting the nominal PWV ($<$0.472)\,mm (Figure \ref{fig:fig3rep}).} 
\label{fig:10000_B910}
\end{figure}

At the longest baselines, configurations C-8 (8.5\,km) to C-10 (16.2\,km)\footnote{Band 9 is limited to the C-8 and C-9 configurations, while Band 10 is limited to the C-8 configuration as of Cycle 9 which offered all configurations. C-9 and C-10 are not offered in Cycle 10}, Bands 3 to 7 also use a 54\,s target scan time, while Band 8, 9 and 10 gradually decrease the target scan times down to 30\,s, i.e. fast switching at Band 10 (see below). There is still $>$85-90\,\% of observing time available at Bands 3 to 5 at night because the phase RMS does not increase significantly with baseline length moving from 5000\,m to 10000\,m (see Section \ref{sec:comparable}). At high frequencies($>$385\,GHz) phase calibrators are weaker and fewer in number \citep[see][]{Asaki2020}. This fact contravenes the requirement of fast cycle times for long baselines, where short calibrator scans are required, because the SNR of calibrators is not always sufficient for successful phase referencing. Only since Cycle 9 have long baseline high frequency observations been offered to the community, in part, as a result of the commissioning of the Band-to-Band (B2B) mode \citep[see][]{Asaki2020} that allows phase calibrators to be observed at a lower frequency band while observing the target at a high frequency band. In using the B2B technique the phase calibrators observed at low frequencies are stronger, meaning that short scan times provide sufficient SNR for phase solutions, and they are more closely located to a target on the sky (see also the ALMA technical handbook\footnote{https://almascience.eso.org/proposing/technical-handbook}). 
Using the Phase RMS database, in Figure \ref{fig:10000_B910} we illustrate the moderate increase in the observing time available from $\sim$5-10\,\% at night time in the winter months to $\sim$10-15\,\% (up to a factor of 1.5 increase dependent on month) at Band 9 and 10 when reducing the timescale from 60\,s to 45\,s, in the left and right panels respectively. We use the phase RMS limit of 30$^{\circ}$ for Band 9 and 40$^{\circ}$ at Band 10, both equivalent to a path length variation of 38\,$\mu$m. The improvement means that the percentage of time available at the 45\,s timescale is constrained by that available when high frequency nominal PWV conditions are met ($<$0.472\,mm), whereas for the 60\,s timescale the percent of observing time available is otherwise limited by the phase RMS requirements and ALMA would not be optimising the use of low PWV conditions should they arise. 

\subsubsection{A note on `Fast-Switching'}  
Fast-switching phase referencing is a technique outlined for use at ALMA since the early Holdaway memos \citep[most recently][]{Holdaway2001}, and was also summarized in \citet{Evans2003}. For ALMA, the fast-switching terminology can be used to describe observations where the regular visits to phase calibrators are made more frequently than $\sim$60\,s\footnote{For perspective, most millimeter-to-radio wavelength interferometers use cycle times of the order minute timescales, as does ALMA for the shortest baselines.}, and hence applies to the longer baseline high frequency observations. The premise, as we have also discussed above, is that we can achieve an overall lower phase RMS for science target scans (bracketed by the phase calibrator) as the atmospheric fluctuations are more frequently calibrated.

In the fast switching simulations by \citet{Holdaway2001} the phase calibrator scans times are reported to be \textit{shorter} than 0.5\,s, while slew times are $<$2\,s. Considering the overall duty cycle (time-on-source / total cycle time) as $\sim$0.9 we infer target scan times of $\sim$40\,s. In their earlier simulation work cycle times of $\sim$20\,s are reported in trying to achieve a 30$^{\circ}$ phase RMS limit for an envisaged 3000\,m baseline configuration \citep[Table 4 in][]{Holdaway1997}. In more recent work \citet{Asaki2014,Asaki2016} tested $\sim$20\,s fast switching at ALMA, although the observations become inefficient with a duty cycle $<$0.5. 

\begin{table}[!t]
\caption{Simple estimates of the duty cycle for phase referencing using both ALMA parameters and hypothetical setups. Slew times are assumed to be 2\,s to, and from, the phase calibrator to target.}
\label{tab:duty}
\begin{center}
\begin{tabular}{l l l l}
\headrow \thead{Target Scan (s)} & \thead{Phase Cal Scan (s)} & \thead{Cycle Time (s)} & \thead{Duty Cycle}  \\
\textbf{54}$^{a}$  &  \textbf{18}  & \textbf{76}  &  \textbf{0.71}  \\ 
54  &  12  & 70  &  0.77  \\
54  &   6  & 64  &  0.84  \\
42  &  18  &  64 &  0.66 \\
\textbf{42}$^{b}$ &  \textbf{12}  &  \textbf{58} & \textbf{0.72} \\
42 &   6  &  52 &  0.81 \\ 
\textbf{30}$^{c}$ &  \textbf{12}  &  \textbf{46} &  \textbf{0.65} \\
30  &  6  & 40  &  0.75  \\
30  &  3$^d$ & 37 & 0.81   \\
18  &  12  &  34 &  0.53 \\
18   &   6  &  28 &  0.64   \\
18  &   3$^d$  & 25 &  0.72  \\ 
\end{tabular}
\\
\end{center}
$^{a}$ setup for ALMA long baselines from Band 3 to 7 to nearest second \\ 
$^{b}$ setup for ALMA long baselines Band 9 to nearest second \\ 
$^{c}$ setup for ALMA long baselines Band 10 to nearest second\\ 
$^d$assuming the integration time is also 3\,s this does not sample the phase calibrator more than once per scan.\\
Note: overall efficiencies (not calculated) would depend on other calibrations as are required for general interferometric observations.
\end{table}


In their long baseline investigations \citet{Maud2022} report that the phase RMS is roughly proportional to $\sqrt{\rm{time}}$, on average, such that a factor of 4 reduction in cycle time (assumed as the target scan time when the phase calibrator scan is considered negligible in length) can improve the phase RMS by $\sim$2, e.g. reducing times from 120\,s down to 30\,s `fast switching'. \citet{Holdaway1997} note that they need to have 6 times more cycles in their simulated phase referencing to improve the phase RMS from 45$^{\circ}$ to 20$^{\circ}$, also roughly following $\sqrt{\rm{time}}$. However, \citet{Maud2022} also note that fast-switching will not necessarily improve all observations, in terms of coherence, when the phase RMS is already good. As we find from the work presented in this memo (above), ALMA can observe in Band 3 for over 90\,\% of the time while meeting the ideal 30$^{\circ}$ phase RMS limit, corresponding to a maximal visibility coherence of 0.87 (Equation \ref{eqn2}) when considering only the atmospheric variability. Thus it is \textit{not necessary} to reduce cycle times by a factor of 4 to reduce the phase RMS to $\sim$15$^{\circ}$ because the coherence improvement is only $\sim$0.09 (to $\sim$0.96) while the time on the target source per cycle would more than half, meaning a considerable drop in efficiency and at least a twice as long total observing time for very little gain (see also Table \ref{tab:duty}). For higher frequencies, where the phase RMS in terms of path length variation needs to be as low as possible, then faster cycle times can notably increase the observing time available (e.g. Figure \ref{fig:10000_B910}). Although the same scenario as above holds for observing efficiency, if for example the phase RMS was $\sim$60$^{\circ}$ (coherence 0.57, Equation \ref{eqn2}), then a reduction to 30$^{\circ}$ using 4 times faster cycle times would provide a 0.3 improvement in coherence (up to 0.87), and would allow previous unsuitable conditions, that would not have provided accurate calibration or images, to be used. The reduction of cycle times from 76\,s to 19\,s however may not be pragmatically possible (see Section \ref{sec:optim}).

For ALMA operations there is a pragmatic limit and phase calibrator scans as short as \citet{Holdaway2001} intended are not really feasible: antennas need to stabilise after moving; the software needs to initiate the scan sequences; data should be recorded for a few integrations - that are typically 3\,s long; if using B2B mode the electronics must also stabilise due to the frequency change; and the duty cycle should not drop so low as to lose efficiency, i.e. when the observing time gained is primarily spent observing the calibrator or slewing. In Table \ref{tab:duty} we estimate the duty cycle for current ALMA long baselines and theoretical phase calibration cycle times. We see that for ALMA's current long baseline parameters the duty cycle is only slightly reduced in the case of Band 10, as compared to standard lower frequency long baselines observing, but the cycle time and target scan times are almost halved. This reduction offers a modest improvement in phase RMS by a factor 1.3 (following $\sqrt{\rm time}$), and as presented above, boosts the time available by a factor $>$1.5. 


\subsection{Forward Look}
\label{sec:optim}
A reduction in cycle times, specifically for the most stringent high frequency observations, can provide notable increases in the observing time available. From the Phase RMS database, using a timescale of 30\,s indicates that the median time available for high frequencies at long baselines can reach $\sim$30\,\%, at night for the winter months, and exceeds what is possible based on the nominal PWV values. For the future a more dynamic, or at least a choice based system could be investigated. One could envisage that the `best' cycle time setup would be selected for the conditions at that specific time, on-the-fly. The cycle time could be reduced for the most `difficult' observations at high frequencies where the PWV is met but the phase RMS might be borderline using a standard setup. The dynamic process could also reduce cycle times for lower frequency observations made during the day time in more unstable conditions, or in reverse use \textit{longer} cycle times that are more efficient if the conditions are very stable. Of course the minimum cycle time would have to be limited as to avoid losing efficiency, but that could easily be compared with a preset pragmatic duty cycle `hard limit'. During the next decade ALMA's Wideband Sensitivity Upgrade (WSU, \citealt{Carpenter2022}) will quadruple the available maximal bandwidth along with providing an improved sensitivity via digital efficiency improvements. In light of these significant upcoming modifications and based on the work presented in this memo, we would recommend a thorough investigation into baseline length and cycle time optimisation, including a fully dynamic system. Ultimately, such optimisations are the way to maximise the science return and also allow more high frequency projects to get on sky. 

For mid-baseline lengths, optimisation is already taking place for ALMA's current high frequency observing capabilities based on the results from the Phase RMS database. The few minute long target scan times will be reduced down to 90\,s in Cycle 10. The Phase RMS database studies showed that going from 180-240\,s timescales down to 90\,s more than doubled the percent of time available in the night time winter months for Bands 9 and 10, while the duty cycle for phase referencing slightly improves as the phase calibrator scan time is also reduced\footnote{In Cycle 9 for high frequency observations the target scan is 180\,s and the phase calibrator scan is 60\,s - duty cycle $\sim$0.75. For Cycle 10, the target scan is 90\,s and the phase calibrator scan is 18\,s - duty cycle $\sim$0.80. However, B2B mode costs an additional overhead \citep[see,][]{Maud2020}.}. Further investigations based on extensions of capabilities work \citep[outlined recently by ][]{Maud2021} that forms part of the ObsMode process \citep[detailed in][]{Takahashi2021} showed that the reduction in phase calibrator scan time from 60\,s to 18\,s did \textit{not} significantly change the average calibrator separation distance from science targets for Band 8, 9 and 10 observations. The close enough and strong enough phase calibrators were still found when using the reduced scan times. Of course the weaker phase calibrators that previously just met the SNR level are no longer suitable, but since the B2B mode is now commissioned for mid-baseline configurations (in Cycle 10), there are \textit{significantly} more calibrators available above the SNR limit when observing them at a lower frequency, such that 18\,s scans are more than sufficient. 

The Phase RMS database will continue to be used to examine more closely individual observations and to explore how the dynamic adjustment of cycle times will affect observing times and efficiencies. 


\section{Summary}
\label{sec:sum}
In this memo we report on ALMA site properties in terms of monthly, daily, hourly variations of the phase RMS based on the examination of the Bandpass source scans of over 17000 observations taken since the start of ALMA Cycle 3. We also reaffirm trends seen for the PWV. In overview we find:
\smallskip
\begin{itemize}
    \item Notable variations of the phase RMS, and PWV, over the year with Chilean summer (December - February/March) providing the worst conditions and winter (June - August) offering the best. 
    \item That night time provides better conditions in offering the most stable phase RMS, and lowest PWV:  
    \begin{itemize}
        \item The diurnal variations in phase RMS are significantly larger than those in PWV.
        \item The median phase RMS (per hour for all months and years) can change by up to a factor of 7 between day and night time in summer months and by up to a factor of 3 in winter months.        
        \item The median PWV fluctuates on average by a factor of 1.9 between night and day averaged over all months. Summer and winter variations are slightly above and below the average respectively.
        \item The average observing time available during the day is $\sim$5-10\,\% lower than at night time for any given nominal PWV/octile level.
    \end{itemize}
    \item The `complete' 5 year sample, in both phase RMS and PWV is not long enough to see any global year to year variations.
\end{itemize}

\bigskip

\noindent Importantly, we report on the variability of the phase RMS on summary baseline lengths of 500, 1000, 5000 and 10000\,m, that are relevant to ALMA array configurations and for timescales closely related to the phase referencing cycle times. Measuring the phase RMS on timescales between 30-240\,s acts as a proxy value of the expected phase errors remaining in any science target scans after phase calibration and hence ties with the expected quality of the final data. We also address bullet point 3 from \citet{Evans2003} in now illustrating the time available for `exciting science' after phase correction would have been applied. We find:  
\smallskip
\begin{itemize}
    \item Application of the Water Vapour Radiometer corrections to the data provides a phase RMS improvement in $>$96\,\% of the observations, and more than a factor of 2 improvement for $\sim$57\,\% of the observations (referenced to the 1000\,m summary baseline and 120\,s timescale).
    \item Observations that do \textit{not} improve with WVR application are mostly comprised of Band 3, or low frequency observations ($<$Band 6), particularly where the PWV is $>$1.5\,mm and the solutions are affected by clouds.    
    \item ALMA's use of faster cycle times for longer baselines can compensate for a naturally increasing phase RMS with baseline length.
    \begin{itemize}
    \item For short baseline, $\leq$1000\,m, \textit{decreasing} the timescales from 240 to 60\,s \textit{increases} the percentage of time that high frequency observations can be made in the available conditions from 18 to 56\,\% at Band 9, using the global cumulative distributions. For $\leq$Band 7 the increase in available time is at most 14\,\% when decreasing from 240 to 60\,s timescales.
    \item For long baselines, $\geq$5000\,m, short timescales $\leq$60\,s are required to allow high frequency observations to be conducted in the conditions available, otherwise the required phase RMS \textit{cannot} be met. 
        \item Typically Band 3 to Band 5 can be observed in all configurations for $>$85-90\,\% of the time at night and $>$70-80\,\% of the time during the day in winter months irrespective of the baseline length when considering ALMA's current observing scheme, using shorter timescales for the longer baseline configurations.
        \item Using the fast switching cycle times for ALMA high frequency long baselines (46\,s) aligns the time available where both the phase RMS and nominal PWV criteria can be met, $>$10-15\,\% of the time available at night for winter months for Bands 9 and 10 observations, when using a nominal PWV of $<$0.472\,mm and $<$38\,$\mu$m path length variation (phase RMS $<$30$^{\circ}$ for Band 9 and $<$40$^{\circ}$ for Band 10).
\end{itemize}
\end{itemize}

\bigskip

\noindent We also find some moderate correlations between weather metadata parameters, such as wind speed, humidity and the PWV and phase RMS. However, the general scatter in parameter space prevents any accurate predictions of the phase RMS. As also concluded by \citet{Evans2003}, we find that low PWV conditions do not necessarily correlate directly to good atmospheric stability (low phase RMS).

 We also suggest how faster, or slower, cycle times could be considered to optimise observations given the observing conditions on a particular day/night where dynamic changes could be made to ensure a low phase RMS was achieved. However, such dynamic cycle time changes requires a deeper study as to trade off any reduction in observing efficiency when using fast-switching compared to the percent of observing time gained, and spent on-source. We recommend a thorough investigation into baseline length and cycle time optimisation, including a fully dynamic system in light of the upcoming Wideband Sensitivity Upgrade.

The Phase RMS database will be available along with helper scripts on the ALMA science portal for public use. Periodic updates are also planning to continue ingesting new cycles of data.

\bigskip
\section*{Appendix}


\subsection*{Database Contents}
\label{sec:db}

\begin{figure}[!h]
\centering
\includegraphics[width=0.8\textwidth]{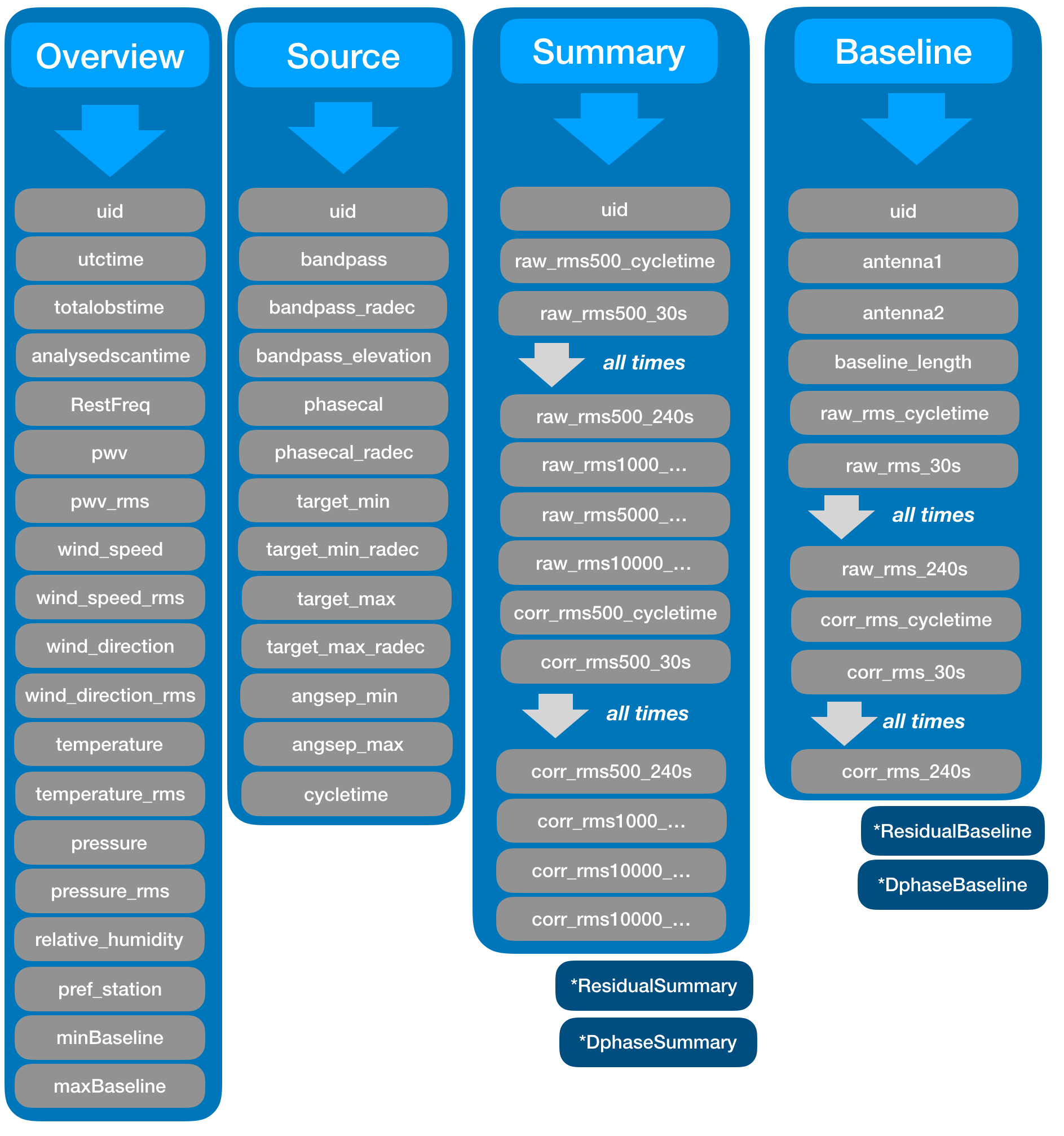}
\caption{Schematic of the Phase RMS SQL database tables and columns. Table names are indicated at the top, and within the respective tables the column headings are shown in grey. Not all columns are listed for the various timescale measures but are noted by \textit{`all times'} and refer to 45, 60, 90, 120, 150, 180, 210\,s times. $^*$the Residual and Dphase Summary and Baseline tables have the same structure as the Summary and Baseline tables.} 
\label{fig:db}
\end{figure}

The Phase RMS database is built with an SQL architecture that is easily portable and can be manipulated with various programming languages, such as {\sc python}. The database structure is not complex and consists of 8 tables in which all EBs are indexed by their unique identification (UID). The tables and a description of their contents are shown in Table \ref{tab:db} with column headings indicated in Figure \ref{fig:db}. The two `metadata' tables are: the \textbf{Source} table containing source information about the Bandpass, Phase calibrator, and the target(s) and their minimum and maximal separation from the phase calibrator and the phase referencing cycle time; and the \textbf{Overview} table storing UTC observing time, total bandpass scan length, wind, temperature, pressure, humidity, PWV and minimum and maximum baseline lengths. The table used throughout this memo is the \textbf{Summary} table that holds the phase RMS values (before and after WVR corrections) as extracted from the phase-time streams for at the various timescales, 30\,s, 45\,s, then 60\,s to  240\, in 30\,s steps and the observing cycle-time\footnote{For observations cycle times exceeding the bandpass scan length the maximal phase RMS value calculated is over the duration of the bandpass scan only.} averaged for baseline lengths of 500, 1000, 5000 and 10000\,m. There is also the \textbf{Baseline} table that stores the phase RMS values per baseline from each observations on the above mentioned timescales. The Baseline table was used to build the Summary table, and in general individual outlier baselines are not ingested into the database.

\begin{table}[!h]
\caption{Total number of EBs in our lists to be ingested into the Phase RMS database}
\label{tab:list}
\centering
\begin{tabular}{l l l r }
\headrow \thead{Year} & \thead{Number} & \thead{Ingested} &  \thead{Failed} \\
2015 & 1889 &  1685  & 204   \\  
2016 & 3611 &  3250 &  357   \\  
2017 & 2732 &  2252  & 480    \\ 
2018 & 4564 & 4054   &  510  \\  
2019 & 3975&  3560  & 415    \\   
\hline
Total & 16771 &  14801 & 1966 \\   


\end{tabular}

\end{table}

At the time of generation (late 2021) for data from 2015 to 2019 inclusive there are 16771 EBs in our list (see Table \ref{tab:list}). As of the date of this memo ingestion of 2020 and 2021 data has begun but is not complete. The database as used for the memo is at a $\sim$87\,\% completeness level with $\sim$2000 EBs that have not been correctly ingested for Cycle 3 (2015) to Cycle 7 (2019). There are a total of 17532 observations ingested including years 2020 and 2021. Note also, there are a few tens of observations with corrupted metadata that were not used in the correlations\footnote{Updated database versions will clean out such observations.}.

\begin{table}[!h]
\caption{Phase RMS database tables and description}
\label{tab:db}
\centering
\begin{tabular}{l l }
\headrow \thead{Table} & \thead{Description}\\
Overview & Observing metadata and weather parameters \\
Source & Source metadata and parameters \\
Summary & Summary baseline length phase RMS values \\
Baseline & All baseline based phase RMS values \\
ResidualSummary & Summary of residual Phase RMS values \\
ResidualBaseline & All baseline based Residual Phase RMS values \\
DphaseSummary & Summary of pseudo D-phase values \\
DphaseBaseline & All baseline based pseudo D-phase values \\

\end{tabular}

\end{table}

Within the database there are also \textbf{Residual} and \textbf{Dphase} versions of the Summary and Bandpass tables also storing the average baseline values and all baseline values, however the phase RMS is calculated differently. For the \textbf{Residual} tables, phase solutions are made for all given timescales by extracting integration times within the Bandpass phase-time stream to act as `calibrator' points, e.g for a 30\,s timescale we take values at 0, 30, 60 ... to $\sim$300\,s for a typical Bandpass scan. A gaintable is produced and applied back to the Bandpass scan data which `corrects' the phases to leave only the \textbf{Residual} variations, for which we then measure the phase RMS from. The \textbf{Dphase} tables present the value from the median difference of the aforementioned solutions, although for timescales $>$90-120\,s the values are not so robust as they can consider only $<$3 points extracted from a typical Bandpass scan, e.g. the 120\,s timescale can use only the median value from absolute phase solutions recorded at 0, 120 and 240\,s. \textit{We store these in the database for future examination but do not discuss them further in the memo}.

\paragraph*{Data Availability:} All materials will be available on the ALMA science portal:\\ https://almascience.eso.org/tools/eu-arc-network. 

\paragraph*{Acknowledgments.} ALMA is a partnership between the European Southern Observatory (ESO), the National Science Foundation (NSF) of the United States and the National Institutes of Natural Sciences (NINS) of Japan in collaboration with the Republic of Chile. ALMA is funded by ESO in representation of its member states, by NSF in collaboration with the National Research Council (NRC) of Canada and the National Science Council (NSC) of Taiwan, and by NINS in collaboration with the Academia Sinica (AS) in Taiwan, and the Korea Astronomy and
Space Science Institute (KASI) of South Korea. 

We are grateful for the discussions with, and hard work of, the Joint ALMA Observatory staff including all of the people who make the observatory `run' and are too many to list.
The authors would particularly like to thank the referees, Sergio Mart\'in, Ruediger Kneissl, Eric Villard, Tony Mroczkowski, and Carlos De Breuck for their helpful and detailed comments, in addition to John Carpenter for his input on an earlier version and George Privon for discussions about figures for the ALMA Proposer's Guide.


\bibliographystyle{aasjournal}

\end{document}